\documentclass[footinbib,amsfonts,floatfix,superscriptaddress,longbibliography,twocolumn]{revtex4-2}
\usepackage[all]{xy}
\usepackage{graphicx}
 \usepackage{mathbbol}
\usepackage{amsmath,amssymb,bm,amsthm}
\usepackage{braket}
\usepackage{soul}
\usepackage[normalem]{ulem}
\usepackage[utf8x]{inputenc}
\usepackage[usenames,dvipsnames, pdftex]{xcolor}
\usepackage{subfigure}
\definecolor{magenta}{rgb}{0.8, 0.0, 0.56}
\usepackage{mathtools}
\usepackage{lmodern}
\usepackage{float}

\begin{document}
	
	\title{From integrability to chaos: the quantum-classical correspondence in a triple well bosonic model}
         \author{Erick R. Castro}
	\email[]{erickc@cbpf.br}
	\affiliation{Centro Brasileiro de Pesquisas F\'{i}sicas/MCTI,
		22290-180, Rio de Janeiro, RJ, Brazil}
	\author{Karin Wittmann W.}
	\email[]{karin.wittmann@ufrgs.br}
	\affiliation{Instituto de F\'{i}sica da UFRGS, Porto Alegre, RS, Brazil}
	\author{Jorge Ch\'{a}vez-Carlos}
	\email[]{jorge.chavez@ciencias.unam.mx}
	\affiliation{Department of Physics, University of Connecticut, Storrs, Connecticut 06269, USA}
	\author{Itzhak Roditi}
	\email[]{roditi@cbpf.br}
	\affiliation{Centro Brasileiro de Pesquisas F\'{i}sicas/MCTI,
		22290-180, Rio de Janeiro, RJ, Brazil}
	\author{Angela Foerster}
	\email[]{angela@if.ufrgs.br}
	\affiliation{Instituto de F\'{i}sica da UFRGS, Porto Alegre, RS, Brazil}
	\author{Jorge G. Hirsch}
	\email[]{hirsch@nucleares.unam.mx}
	\affiliation{Instituto de Ciencias Nucleares, Universidad Nacional Aut\'{o}noma de M\'{e}xico,
		Apdo. Postal 70-543, C.P. 04510 Cd. Mx., Mexico}

\begin{abstract}
In this work, we investigate the semiclassical limit of a simple bosonic quantum many-body system exhibiting both integrable and chaotic behavior. A classical Hamiltonian is derived using coherent states. The transition from regularity to chaos in classical dynamics is visualized through Poincaré sections. Classical trajectories in phase space closely resemble the projections of the Husimi functions of eigenstates with similar energy, even in chaotic cases. It is demonstrated that this correlation is more evident when projecting the eigenstates onto the Fock states. The analysis is carried out at a critical energy where the eigenstates are maximally delocalized in the Fock basis. Despite the imperfect delocalization, its influence is present in the classical and quantum properties under investigation. The study systematically establishes quantum-classical correspondence for a bosonic many-body system with more than two wells, even within the chaotic region.  

\end{abstract}

	\maketitle
	
	\section{Introduction}
	As is well known, the field of quantum chaos begins by identifying singular features present in quantum systems, obtained through the quantization of chaotic classical systems \cite{FritzHaake, Gutzwiller, Gut1971, Aurich1994, Boh1993}. While classical chaos is uniquely associated with the explicit non-linear character of the equation of motion, the term ``quantum chaos" refers specifically to the system's behavior in the quantum regime. The most extensively studied example of this classical-quantum correspondence is the billiard system, where precise and unambiguous relations have been established \cite{BGC1984, Seba1990, Lozej2022}. When a chaotic system is quantized, the separation between its eigenvalues exhibits a Wigner-Dyson distribution, representing a typical signature for recognizing quantum chaos. Another paradigmatic example with a well-established chaotic quantum-classical correspondence is the Dicke model \cite{RHD1954}, and its study is a highly active research field \cite{BAF2013, PVBLSH2021, Emary2003, Magnani2015, Wang2022}.

	Other signatures are also employed to characterize standard features of chaotic quantum systems, including correlation holes, a Gaussian distribution of the density of states, and an off-diagonal distribution in the quantum evolution of mean values of observables \cite{Gorin2002, Herrera2017, Santos2012, French1970, Brody1981, FR2012, Mondaini2017, Santos2020}. These distinctive features are also present in quantum systems with no explicit classical limit (such as nuclear or many-body quantum systems) and in general random matrix models, where these signatures are understood as typical universal behavior in interacting non-integrable quantum systems.
		
	This work analyzes the quantum-classical correspondence of a three-well bosonic model introduced in \cite{WYTLA2018} and further studied in \cite{WCFS2022} from a pure quantum chaos perspective. The model can be considered a generalized Bose-Hubbard model with an integrable limit and exhibits mixed behavior in the chaotic regime, as asserted by other recent studies of the Bose-Hubbard model \cite{NH2022}. Perturbed integrable models with chaotic limits are intriguing mathematical artifacts, as they provide the opportunity for a detailed study of the integrable-chaotic transition, shedding light on the intricate aspects of the mixing process. The studied model belongs to a family of integrable bipartite models \cite{Ymai2017}, \footnote{For the four-well case, see ~\cite{grun2022interfer} and ~\cite{grun2022protocol}, where the integrable properties were studied to produce interferometry and NOON states.}, descending from particular two-well bosonic models. Integrability is associated with nontrivial intrinsic symmetries determined by the Bethe ansatz. Here, we thoroughly explore the quantum-classical correspondence in this model for the integrable, mixed, and chaotic regimes.

	The Bose-Hubbard model is intrinsically connected with the vast field of Bose-Einstein condensates. It represents a specific limit of the spatial quantum wave function within certain interacting potentials, describing the behavior of ultracold atoms and diluted alkaline gases \cite{Legget2001, Bloch2008}. Extensive literature covers various aspects of this significant discipline, including localization, catastrophe theory, quantum chaos, superfluidity, and the generation and control of entanglement
\cite{Viscondi2011,Cao2011,Koutentakis2017,Bloch2005,Fisher1989,Kollath2010,Dutta2019,Oelkers2007,Guo2014,Kolovsky2004,Choy1982,Streltsov2011,Lahaye2010,Buonsante2009,Foerster2007,tonel2020SciPost,Nakerst2021,Rautenberg2020,Ray2020,Bera2019,Richaud2018,Garcia_March_2018,Guo2018,McCormack2021,Yan2023,Dag2023,wittmann2023}.

	The paper is structured as follows: Section \ref{section1} defines the quantum Hamiltonian, and we determine the classical critical energy $E_{\mathrm{Classic}}$ through a pure eigenstates analysis. The nearest eigenstates to $E_{\mathrm{Classic}}$ are also the more delocalized eigenstates, and we select this energy region for our study. Subsequently, we deduce the classical Hamiltonian written in an appropriate set of coordinates and establish interesting properties in the classical trajectories. Section \ref{sectionIII} introduces coherent states and, from them, the Husimi function, which measures quantum localization in phase space. The set of Fock states can also be used for the same purpose. Remarkably, we find correspondence for many classical short-time trajectories and quantum localization in phase space for many eigenstates in the chaotic regime. Specific patterns and configurations are analyzed.

 	In Section \ref{section2}, we analyze the classical quantum correspondence of certain trajectories when evolved over long periods of time. The study is conducted in three regimes: regular (integrable), mixed, and chaotic. In the regular regime, we use the associated symmetry to establish the correspondence in a precise form. In the chaotic case, the correspondence is established using microcanonical mean values of the Husimi functions, and we show that the classical pattern obtained at large times is found in the quantum case because remnant localization persists in the set of eigenstates. This is expressed as a slight deviation from the Gaussian distribution of the components.
	
	Section \ref{section5} contains the final discussion of the results obtained in this work.

\section{Quantum and classical Hamiltonians and the critical energy}\label{section1}
\subsection{Quantum Hamiltonian}	
	The Hamiltonian of our model is an extension of a two-well potential version studied in \cite{TLF2005, Links2006}. It describes $N$ bosons in an aligned three-well potential and is given by
	\begin{align}
	\label{QH}
	\hat{H} =& \frac{U}{N}\left(\hat{N}_1-\hat{N}_2+\hat{N}_3\right)^2 + \epsilon\left(\hat{N}_3-\hat{N}_1\right) \nonumber \\ &+\frac{J}{\sqrt{2}}\left(\hat{a}_1^\dagger \hat{a}_2 + \hat{a}_2^\dagger \hat{a}_1\right)+\frac{J}{\sqrt{2}}\left(\hat{a}_2^\dagger \hat{a}_3 + \hat{a}_3^\dagger \hat{a}_2\right),
	\end{align}
   where $\hat{N}_k=\hat{a}_k^{\dagger}\hat{a}_k$ is the number operator of the $k$th-well, $\hat{a}_k^{\dagger}$ and $\hat{a}_k$ are respectively the creation and annihilation operators of the $k$th-well. The total particle number $N=N_1+N_2+N_3$ is fixed. $U$ is the magnitude of the boson interactions, $J$  parameterize the jump between wells, and $\epsilon$ is an external tilt. If $\epsilon=0$, the system is invariant by the interchange of wells 1 and 3 and is integrable; if $\epsilon\neq 0$, the integrability is broken.  

   The Hamiltonian (\ref{QH}) can be expressed in a matrix representation through occupation number (Fock) states $|N_1,N_2,N_3\rangle$. The associated Hamiltonian matrix has dimension $D\times D$ with $D=\frac{(N+2)!}{2!N!}$, and its $D$ eigenvalues are obtained by numerical diagonalization.
	
The spectrum of this quantum Hamiltonian, for different parameter values, exhibits both regularity and chaos regions. Various cases were analyzed in the integrable limit $\epsilon=0$: the Rabi and Fock limits, which are highly degenerate, and the Josephson regime in between. For $\epsilon > 0$, a significant number of avoided crossings are observed, and at $\epsilon \approx 1.5$, the spectrum displays the characteristic level repulsion of quantum chaotic systems \cite{WCFS2022}.

\subsection{Classical critical energy and Shannon entropy}
    
    A comprehensive calculation of the equilibrium points associated with the classical limit of the Hamiltonian \eqref{QH} was conducted in \cite{CCRSH2021}. The stable configurations with minimal and maximal energy were identified, showing a complete correspondence with the minimal and maximal eigenvalues for sufficiently large $N$. While these results focus on ground-state properties, the tools employed can also be utilized to compute intermediate energy configurations, which, in principle, are associated with possible excited quantum phase transitions. 
   These intermediate configurations are consistently present, particularly in the vicinity of the chaotic regime identified by the parameters  $(U, J,\epsilon)=(0.7, 1, 1.5)$. In this configuration, an unstable critical point arises with energy $E_{\mathrm{Classic}} \approx 0.0752$, represented by the normalized state 
   $$\mathbf{N} = (N_1/N,N_3/N,\phi_ {12},\phi_{32}) \approx(0.081, 0.294, 0, \pi),$$ 
   where $\phi_ {12}$ and $\phi_{32}$ represent the phases in the classical Hamiltonian approach, as detailed in the next subsection.
    
\begin{figure}
		\centering
   	 	\includegraphics[width=1\linewidth]{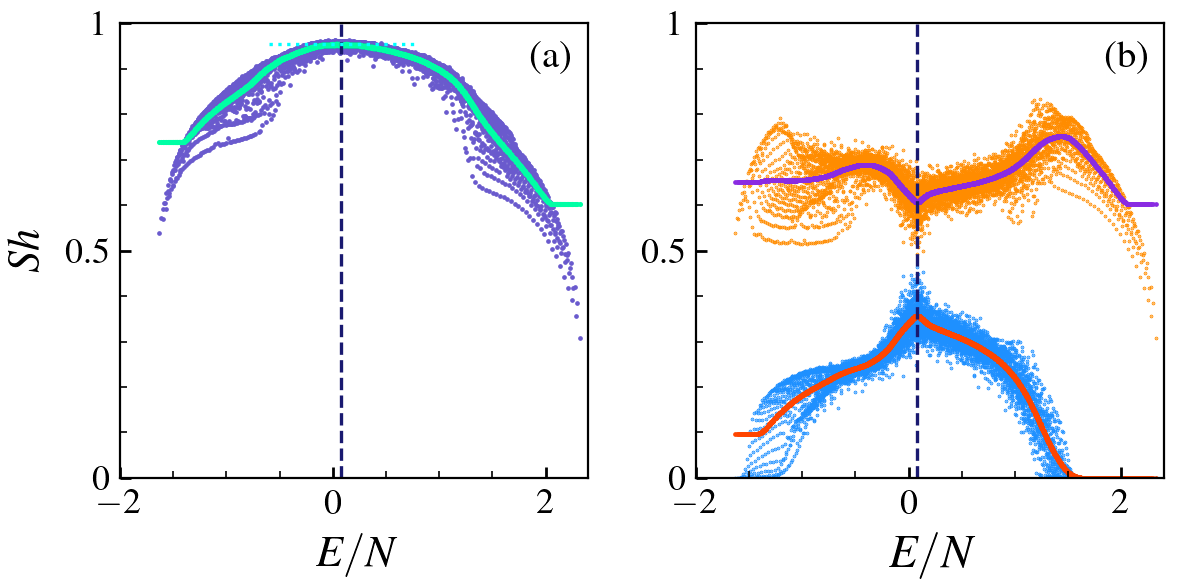}
		\caption{We can express the total Shannon entropy as $Sh_{\mathrm{Total}}^{E}=Sh_{N_2\geq N_1+N_3}^{E}+Sh_{N_2 < N_1+N_3}^{E}$; the total Shannon entropy is depicted in purple in (a). In (b), the blue and orange figures correspond, respectively, to $Sh_{N_2\geq N_1+N_3}^{E}$ and $Sh_{N_2 < N_1+N_3}^{E}$. The dashed blue line corresponds to the critical energy, and the curved lines represent the mean value of the Shannon entropy around the 200 eigenstates closest to each $E$. We use $\epsilon=1.5$ and $N=120$.}
		\label{fig07}
	\end{figure}

    It has been found that the presence of classical critical points can be reflected in the structure of quantum eigenstates. For instance, it was demonstrated that the participation ratio is sensitive to the existence of the critical point when considering only states in a basis with a particular symmetry \cite{STB2016}. Regarding our model, the Shannon entropy can be examined for each eigenstate in the Fock basis, and no evidence of abrupt changes corresponding to the critical classical energy has been observed \cite{WCFS2022}. In Fig. \ref{fig07}(a), the Shannon entropy of the eigenstates is plotted against the eigenenergy. The curved light blue line represents the mean value of the Shannon entropy around the 200 eigenstates closest to each $E$. No singular patterns are observed around the critical energy, marked with a vertical line.
		
    We observe that the critical point with classical energy $E_{\mathrm{Classic}}$ satisfies $N_2>N_1+N_3$ (From $\mathbf{N}$, we see that $N_2/N \approx 0.625$). Considering only the Fock states with $N_2> N_1+N_3$ (or equivalently with $N_2<N_1+N_3$), we can determine the critical point energy using the Shannon entropy restricted to this particular set. In Fig.\ref{fig07}(b) it is shown an appreciable change in $E/N \approx 0.0752$ when the Shannon entropy is restricted to the particular set of Fock states that satisfy $N_2> N_1+N_3$ (orange points for $N_2<N_1+N_3$).

As the total Shannon entropy reaches its maximum value at this point (see Figure \ref{fig07}a), the components of the related eigenstates are more delocalized; therefore, they are suitable for quantum-classical correspondence analysis in the chaotic regime.
For the analysis of the quantum chaotic regime, the parameters $(U, J, \epsilon) = (0.7, 1, 1.5)$ will be adopted, and as a reference, the energy $E_{\mathrm{Classic}}$. In quantum chaos, eigenstates with eigenvalues $E/N$ closest to $E_{\mathrm{Classic}}$ will be considered. For consistency, states of the mixed and regular (integrable) regimes will also be analyzed using the same $(U, J)$ and energy parameters.
   
\subsection{Classical Hamiltonian and dynamics}
	The classical Hamiltonian is obtained using coherent states, $|\alpha\rangle=|\alpha_1,\alpha_2,\alpha_3 \rangle$, where ${\alpha_k=\sqrt{N_k}\exp(i\phi_k)}$. It leads to
	\begin{align}
	\label{CH1}
	\mathcal{H}_{\text{cl}} &= \frac{\langle \alpha |\hat{H} |\alpha \rangle}{N}   \\ &= \frac{U}{N}\left(N_1-N_2+N_3\right)^2 + \epsilon\left(N_3-N_1\right) \nonumber \\ + &J \sqrt{2} \left[\sqrt{N_1N_2}\cos(\phi_1-\phi_2)+\sqrt{N_2N_3}\cos(\phi_2-\phi_3)\right]. \nonumber 
	\end{align}
	
	Defining $\rho_k=\sqrt{N_k/N}$, the classical Hamiltonian reads
	\begin{align}
	\label{CH2}
	\bar{\mathcal{H}}_{\text{cl}} =&	\frac{\mathcal{H_{\text{cl}}}}{N} =
	U\left(\rho_1^2-\rho_2^2+\rho_3^2\right)^2 + \epsilon\left(\rho_3^2-\rho_1^2\right) \nonumber \\ 
	+&J \sqrt{2} \left[\rho_1\rho_2\cos(\phi_1-\phi_2)+\rho_2\rho_3\cos(\phi_2-\phi_3)\right]  \\ 
	& = \frac{U}{4} \left(Q_1^2+P_1^2-Q_2^2-P_2^2+Q_3^2+P_3^2\right)^2 \nonumber \\ 
	&+ \frac{\epsilon}{2}\left(Q_3^2+P_3^2-Q_1^2-P_1^2\right) \nonumber \\ 
	&+\frac{J}{ \sqrt{2}} \left[Q_1 Q_2 + P_1 P_2 + Q_2 Q_3 + P_2 P_3\right]\label{CH3}
	\end{align}
	In the last line, we have introduced the canonical variables
	\begin{align}
	Q_1 = \sqrt{2}\rho_1 \cos(\phi_{1}), \,\,& P_1 = \sqrt{2}\rho_1 \sin(\phi_{1}), \\
	Q_2 = \sqrt{2}\rho_2 \cos(\phi_{2}), \,\,& P_2 = \sqrt{2}\rho_2 \sin(\phi_{2}), \\
	Q_3 = \sqrt{2}\rho_3 \cos(\phi_{3}), \,\,& P_3 = \sqrt{2}\rho_3 \sin(\phi_{3})
	\end{align}
	
	How many degrees of freedom has the above classical Hamiltonian? There are 3 generalized coordinates $Q_1,Q_2,Q_3$ and their conjugated momenta $P_1,P_2,P_3$. 
	There are two known conserved quantities: the energy $E$, given that the Hamiltonian is time independent, and the total number of bosons $N=N_1+N_2+N_3$, implying that  $\rho_1^2+\rho_2^2+\rho_3^2 = 1$.
	
	It can also be seen in Eq. (\ref{QH}) and (\ref{CH2}) that the Hamiltonian is also invariant under the continuous transformation $\phi_k \rightarrow \phi_k+ \theta$, for an arbitrary phase $\theta$.
 
	\begin{align}\label{Transf1}
	Q'_k =& \sqrt{2}\rho_k \cos(\phi_{k} +\theta)   \nonumber \\
	=& \sqrt{2}\rho_k (\cos(\phi_k) \cos(\theta)-\sin(\phi_k) \sin(\theta)) , \nonumber \\
	=& Q_k \cos(\theta)- P_k \sin(\theta) ,
 	\end{align}
        \begin{align}\label{Transf2}
	\,\,P'_k =& \sqrt{2}\rho_k \sin(\phi_{k} + \theta), \nonumber \\
	=& \sqrt{2}\rho_k(\sin(\phi_k) \cos(\theta)+\cos(\phi_k) \sin(\theta)) , \nonumber \\
	=&P_k \cos(\theta)+ Q_k \sin(\theta)) 
	\end{align}
	
	It follows that $Q_k^2-P_k^2 = {Q'_k}^2 - {P'_k}^2 $. Also
	\begin{align}
	Q'_1 Q'_2 + P'_1 P'_2 &= Q_1 Q_2 + P_1 P_2 
	\end{align}
 
 Therefore, the Hamiltonian is invariant under this rotation in each phase space by the same angle $\theta$. The trajectories $\{Q_k(t),P_k(t) \}$, obtained as solutions of the dynamical Hamilton equations, will be described by the same mathematical expressions in the different coordinate systems. It implies that the trajectories with initial conditions $\{Q'_1(\theta),Q'_2(\theta),Q'_3(\theta),P'_1(\theta),P'_2(\theta),P'_3(\theta)\}$ can be obtained from those with initial conditions  $\{Q_{1}(0),Q_{2}(0),Q_{3}(0),P_{1}(0),P_{2}(0),P_{3}(0)\}$ by applying the transformation given by Eqs. \eqref{Transf1} and \eqref{Transf2}.
    
	With the above tools, we can consider expressing the Hamiltonian employing the dynamical variables of the classical system, which are the $\rho_k$'s and the phase differences $\phi_{k,k+1}=\phi_k-\phi_{k+1}$.  Noticing that $\rho_2 = \sqrt{1 - \rho_1^2 -\rho_3^2}$ and that only the differences in phases are relevant in the Hamiltonian, we can define $\cos(\phi_1-\phi_2) = \cos(\phi_{12})$ and  $\cos(\phi_2-\phi_3) = \cos(\phi_{23})$.

	The Hamiltonian takes the simplified form
	\begin{align}\label{Hrho}
	\bar{\mathcal{H}}_{\text{cl}} =& U\left( 2(\rho_1^2+\rho_3^2)-1\right)^2 + \epsilon\left(\rho_3^2-\rho_1^2\right) \nonumber \\ 
	+&J \sqrt{2}  \sqrt{1 - \rho_1^2 -\rho_3^2} \left[\rho_1 \cos(\phi_{12})+ \rho_3 \cos(\phi_{23})\right] \end{align}
	This Hamiltonian has 4 classical variables  $\rho_1, \phi_{12}$, $\rho_3, \phi_{23}$.
	We can introduce the canonical variables 
	\begin{align}\label{T1}
	q_1 = \sqrt{2}\rho_1 \cos(\phi_{12}), \,\,& p_1 = \sqrt{2}\rho_1 \sin(\phi_{12}), \\
	q_3 = \sqrt{2}\rho_3 \cos(\phi_{23}), \,\,& p_3 = \sqrt{2}\rho_3 \sin(\phi_{23})\label{T2}.
	\end{align}
	In terms of these new variables, the Hamiltonian reads
	\begin{align}\label{CH4}
	\bar{\mathcal{H}}_{\text{cl}} = U \left( q_1^2+p_1^2+q_3^2 + p_3^2-1\right)^2  \nonumber \\ 
	+ \frac{\epsilon}{2}\left(q_3^2 + p_3^2-q_1^2-p_1^2\right) \nonumber \\ 
	+J(q_1+ q_3)\sqrt{1 - \frac{(q_1^2+p_1^2+q_3^2 + p_3^2)}{2}}. \end{align}
	
	The classical Hamiltonian (\ref{CH3}) has three classical degrees of freedom and two conserved quantities: $E$ and $N$. Hamiltonian (\ref{CH4}) has two degrees of freedom and only one conserved quantity, the energy $E$. Both are equivalent and display chaos, as we will show later.
 
 The Hamilton equations \eqref{CH2} have the form $$\left(\dot{Q}_i,\dot{P}_i\right)=\left(\frac{\partial 	\bar{\mathcal{H}}_{\text{cl}}}{\partial P_i},-\frac{\partial 	\bar{\mathcal{H}}_{\text{cl}}}{\partial Q_i}\right)=\left(f_i,g_i\right)$$ for $i\in\{1,2,3\}$. In Eq. \eqref{CH2},  the $f_i$ and $g_i$ are simple polynomials in the three pair $(Q_k,P_k)$'s of variables. However, in Eq. \eqref{CH4} the functions $f_i$ and $g_i$ in the two pairs $(q_k,p_k)$'s are more involved. It implies that in some cases, it is numerically more efficient to employ Eq. \eqref{CH2} to calculate the classical trajectories using the Julia free software \textit{Taylor.Integration} \cite{HernandezPerez2019}.

 The classical Hamiltonian depends on four variables. Throughout this work, we choose $\rho_2=0$ as one of the initial conditions, this particular configuration exhibits interesting properties, and we will conclude this section by studying it.

The $\rho_2=0$ condition implies that $\rho_1^2+\rho_3^2=1$, and the $J$ 
 term in Eq.\eqref{Hrho} is zero, indicating that the classical energy is independent of the variables $\phi_{12}$ and $\phi_{32}$.
 This condition arises when $\epsilon=0$ only for configurations with classical energy $E=U$. 
 For $\epsilon>0$, the condition occurs for a unique pair $(N_1,N_3)$ when the energy is fixed. In $\rho_2=0$, all configurations with arbitrary phases $\phi_{12}$ and $\phi_{32}$ have the same energy.
 
\begin{figure}
    \centering
      	 	\includegraphics[width=1\linewidth]{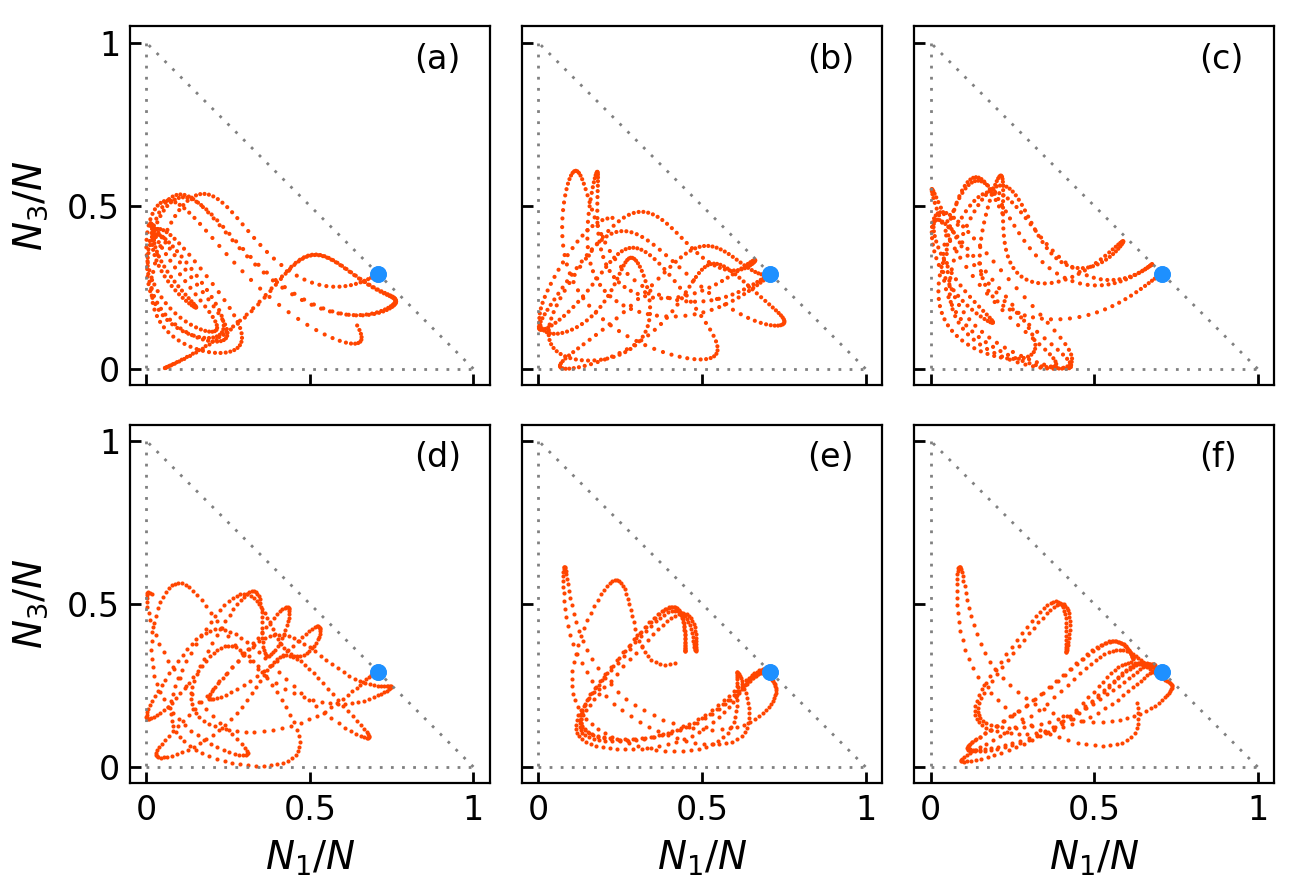}
\caption{Classical trajectories for $E_{\mathrm{Classic}} \approx 0.0752$ and $(U, J, \epsilon)=(0.7, 1, 1.5)$ with an initial condition at $N_2=0$ (blue point) and phase differences $\phi_{32}-\phi_{12} \in {0, \pi/4, \pi/2, 3\pi/4, \pi, 5\pi/4}$, where $\phi_{12}=-\pi$. The trajectories are distinct and originate from the same $(N_1/N, N_3/N)=(0.7082, 0.2917)$ value. This is the only coordinate pair where $N_1+N_3=N$ ($N_2=0$), which is possible for $E_{\mathrm{Classic}}$.}
		\label{fig00A}
	\end{figure}

In Fig. \ref{fig00A}, for $\epsilon=1.5$, six trajectories are shown with the initial condition $\rho_2=0$, for the unique pair $(N_1,N_3)$, represented as a large blue dot. Such initial conditions have different phase differences $\phi_{32} - \phi_{12}$, generating distinct trajectories with the same energy as the figure shows.
 
\begin{figure}
		\centering
   	 	\includegraphics[width=1\linewidth]{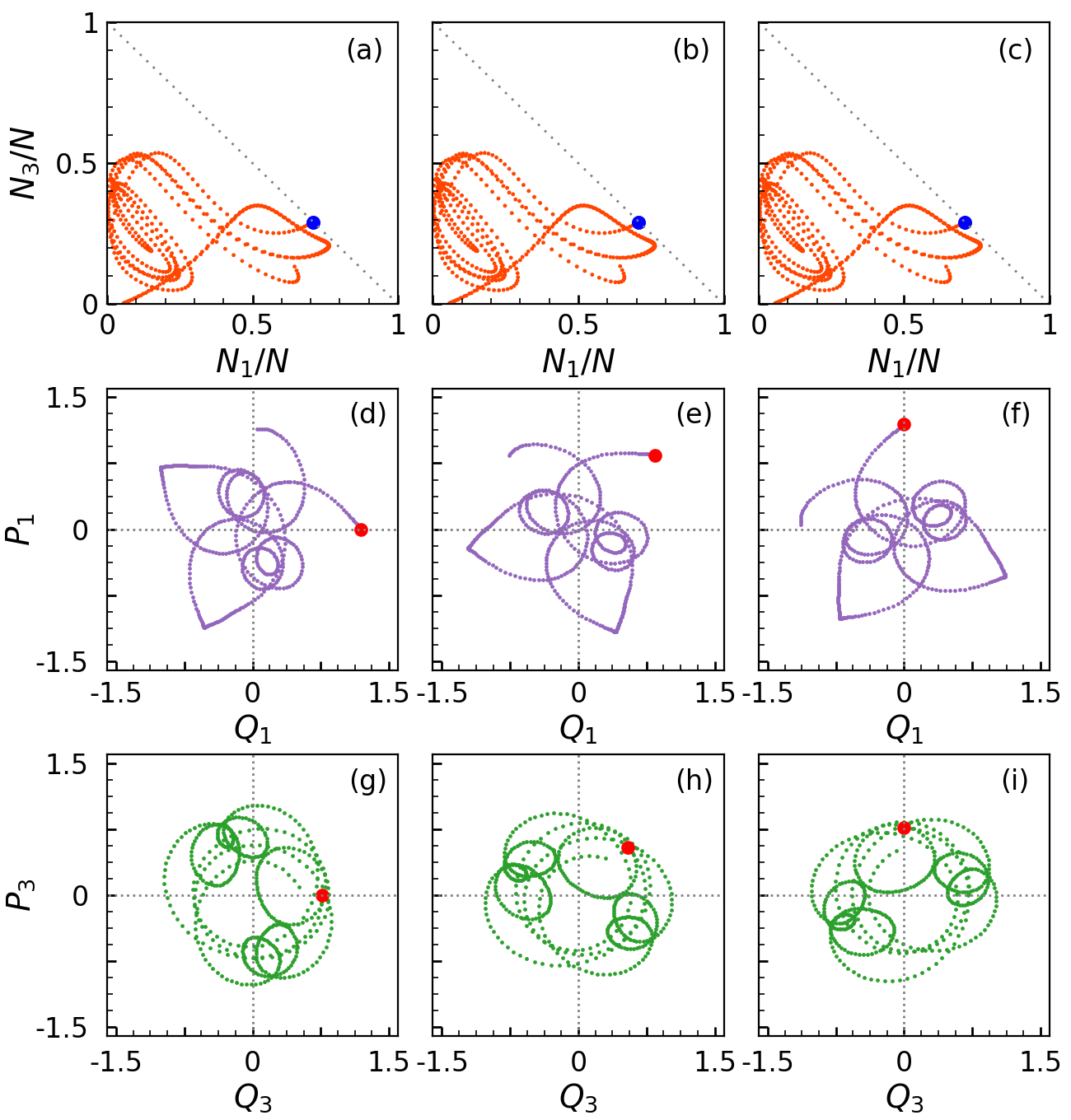}
		\caption{Classical trajectories for $E_{\mathrm{Classic}}\approx 0.0752$ and $(U, J, \epsilon)=(0.7, 1, 1.5)$ with the initial condition at $N_2=0$ (blue point) and an equal phase difference $\phi_{32}-\phi_{12} = 0$, with values $\phi_{12}=\phi_{32}=0$, $\pi/4$, and $\pi/2$. The trajectories in the $\left(N_1, N_3, \phi_{12}, \phi_{32}\right)(t)$ coordinates are identical (top figures a, b, and c), and the mutually equal changes in $(\phi_{12}, \phi_{32})$ rotate the trajectory in the $(Q_1, P_1)$ space (middle figures) and the $(Q_3, P_3)$ space (bottom figures).}.
		\label{fig00B}
	\end{figure}

Additionally, for $N_2=0$, different choices of phases $(\phi_{12},\phi_{32})$ with an identical phase difference $\phi_{32}-\phi_{12}$ generate the same trajectory. In this case, it is possible to assume that $\phi_2=0$ and therefore $\phi_{12} = \phi_1$ and $\phi_{32} = \phi_3$. While these two phases are rotated by the same angle $\theta$ to maintain their difference, the situation differs from Eqs. \eqref{Transf1}-\eqref{Transf2}, because the phase $\phi_2$ is held constant (and is null). In this case,  the same trajectories are obtained but rotated by an angle $\theta$ in the planes $(Q_1,P_1)$ and $(Q_3,P_3)$. This effect can be observed in Fig. \ref{fig00B}, where three sets of phases $\phi_{32},\phi_{12}$ are rotated by the same angle. In the plane $(N_1,N_3)$, the trajectory remains the same, while in the planes $(Q_1,P_1)$ and $(Q_3,P_3)$, the trajectories have the same shape but are rotated.

\section{Quantum localization in phase space}\label{sectionIII}
While classical trajectories cannot be directly mapped to quantum states due to the uncertainty principle, the projection of a quantum state onto coherent states allows us to obtain the Husimi quasi-probability distribution in phase space.

    The coherent state $|\alpha_1,\alpha_2,\alpha_3\rangle$ includes all possible total numbers of bosons in the three wells. As the system under study has a fixed total number of bosons, it is useful to introduce a coherent state projected to a fixed total particle number, expressed as:
    \begin{align}\label{CohState}
    |N_1,&N_3,\phi_{12},\phi_{32} \rangle=\nonumber \\
    &\underset{n_1+n_2+n_3=N}{\sum_{n_1,n_2,n_3}}\left[P\left(n_1,n_2,n_3,\frac{N_1}{N},\frac{N_2}{N},\frac{N_3}{N}\right)\right]^{1/2}\nonumber \\
    &\,\,\,\,\,\,\,\,\,\,\,\,\,\,\,\,\,\,\,\,\,\,\,\,\,\,\,\,\,\,\times e^{in_1\phi_{12}}e^{in_3\phi_{32}}|n_1,n_2,n_3\rangle
    \end{align}
    where $P$ is the multinomial distribution 
    \begin{equation}\label{Multi}
    P\left(n_1,n_2,n_3,p_1,p_2,p_3\right)=\frac{N!}{n_1!n_2!n_3!}p_1^{n_1}p_2^{n_2}p_3^{n_3},
    \end{equation}
    $\phi_{jk}=\phi_j-\phi_k$ and $p_1+p_2+p_3=1$. Note that each classical initial condition $(N_1, N_3,\phi_{12},\phi_{32})$ is associated with the coherent state $|N_1, N_3,\phi_{12},\phi_{32} \rangle$.
    
    The projection of the Husimi function of an eigenstate $|E\rangle$ with eigenenergy $E$ over the plane $(N_1, N_3)$ can be calculated using	
	\begin{equation}\label{Eq:Husimi}
	\mathfrak{H}_{E}^{\mathcal{T}}(N_1,N_3)=\int_{-\pi}^{\pi}\int_{-\pi}^{\pi}\frac{d\phi_{12}}      {2\pi}\frac{d\phi_{32}}{2\pi}\left|\langle E|N_1,N_3,\phi_{12},\phi_{32}\rangle\right|^2
	\end{equation}
 
    An alternative description of the localization of the quantum states in the same plane $(N_1,N_3)$ of the phase space is provided by the projection of an eigenstate $|E\rangle$ on the Fock states 
    \begin{equation}\label{HusFock}
    \mathfrak{H}_E\left(N_1,N_3\right)=\vert \langle N_1,N_2= N-N_1-N_3,N_3 \vert E \rangle \vert^2.
    \end{equation}

    The principal difference is that the quantity $\mathfrak{H}_E\left(N_1,N_3\right)$ generally exhibits a strong interference pattern in non-chaotic regimes, with the advantage that the computational numerical cost is very economical compared to the Husimi function. Fock states have also been utilized in reference \cite{Kirkby2022} to compare quantum-classical correspondence, focusing on the quantum mean value evolution in Fock states and caustics for the classical system at different times.

    \begin{figure}
        \centering
                \includegraphics[width=0.95\linewidth]{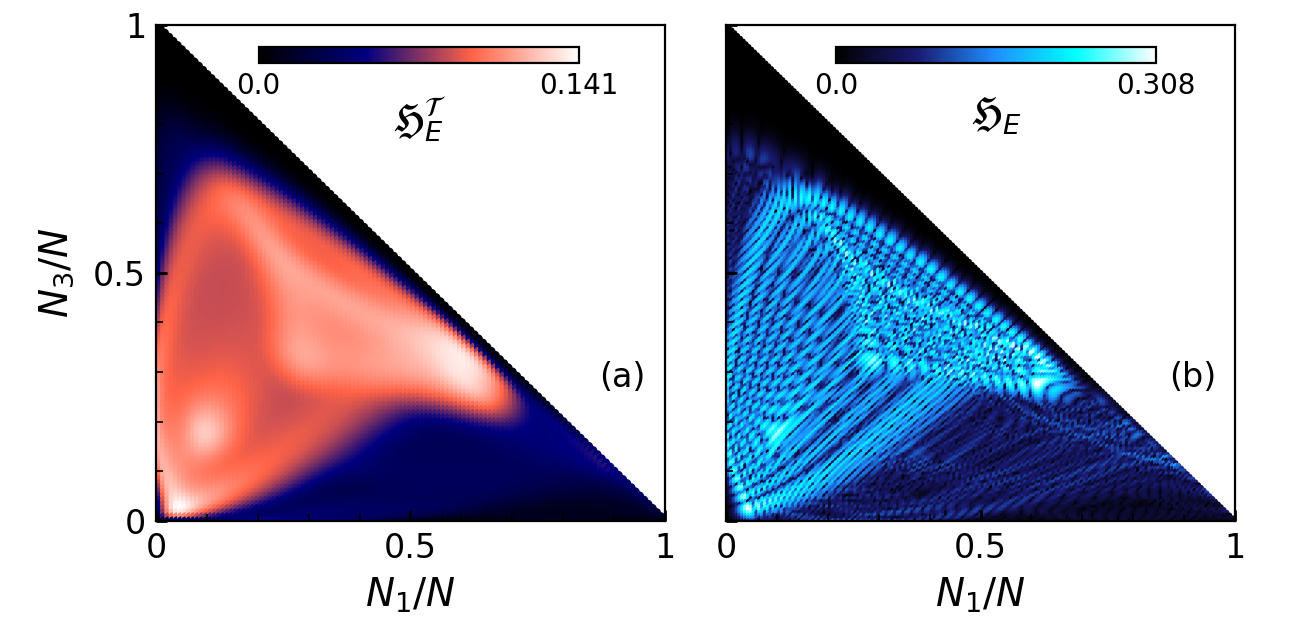}
        \caption{Projected Husimi function $\mathfrak{H}_{E}^{\mathcal{T}}(N_1,N_3)$ (a) and the function $\mathfrak{H}_{E}(N_1,N_3)$ using Fock states (b), for $E \approx E_{\mathrm{Classic}}$,  $\epsilon=0.7$ and $N=180$. We exponentiate each point by 0.25 in both figures to obtain more details.}
        \label{fig01}
    \end{figure}

    In Fig. \ref{fig01}, the two functions $\mathfrak{H}_{E}^{\mathcal{T}}(N_1,N_3)$ and $\mathfrak{H}_{E}(N_1, N_3)$ are compared for an eigenstate with energy close to $E_{\mathrm{Classic}}$, $N=150$, $\epsilon=0.7$, $U=0.7$ and $J=1$, of the mixed regime.
    Note the similarity between the two figures, indicating that in this particular projection, the function $\mathfrak{H}_{E}(N_1, N_3)$ (which utilizes Fock states) provides similar but sharper information than the Husimi function. In particular, the function $\mathfrak{H}_{E}(N_1,N_3)$ in Fig. \ref{fig01}(b), exhibits a strong interference pattern, offering better resolution than the Husimi function, with the additional advantage of being very efficient in numerical computational terms compared to the standard Husimi function.

    To understand the similarity between $\mathfrak{H}_{E}^{\mathcal{T}}$ and $\mathfrak{H}_{E}$, we can introduce Eq. \eqref{HusFock} into Eq. \eqref{Eq:Husimi}, obtaining

    \begin{align}
    \mathfrak{H}_{E}^{\mathcal{T}}\left(N_1,N_3\right)=\nonumber \\
    &\underset{n_1+n_2+n_3=N}{\sum_{n_1,n_2,n_3}}P\left(n_1,n_2,n_3,\frac{N_1}{N},\frac{N_2}{N},\frac{N_3}{N}\right)\nonumber \\
    &\,\,\,\,\,\,\,\,\,\,\,\,\,\,\,\,\,\,\,\,\,\,\,\,\,\,\,\,\,\,\,\,\,\,\,\,\times \mathfrak{H}_{E}\left(n_1,n_3\right).
    \end{align}
    We can observe that the Husimi Function $\mathfrak{H}_{E}^{\mathcal{T}}(N_1,N_3)$ is essentially the eigenstate projection $\mathfrak{H}_{E}(n_1,n_3)$ averaged over all Fock states weighted by the distribution $P(n_1,n_3,\frac{N_1}{N},\frac{N_3}{N})$. The distribution is peaked at $n_i=N_i$ for all $i$, and for large $N$, $\mathfrak{H}_{E}(n_1 \approx N_1,n_3 \approx N_3)$ provides the primary contribution. This results in the observed similarity.


\subsection{Classical trajectories vs quantum projections on phase space}
\begin{figure*}
	 	\centering
             	 	\includegraphics[width=0.98\linewidth]{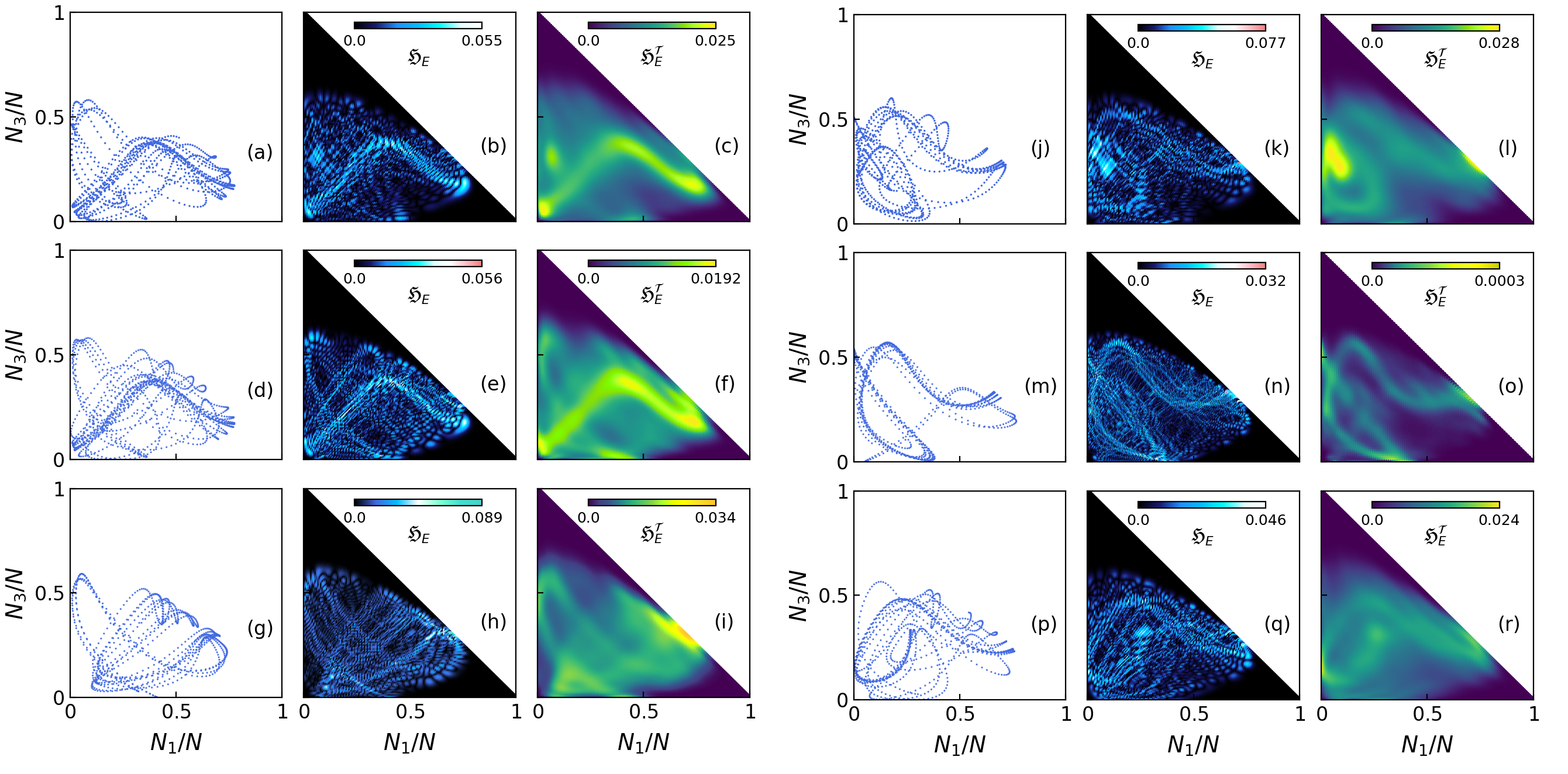}\quad
         	 	\caption{Classical trajectories projected onto the $(N_1,N_3)$ plane for six distinct initial conditions with energy $E_{\mathrm{Classic}}$ (a, d, g, j, m, p). Equivalent projections are obtained for the Husimi Function for particular eigenstates with energy $E \approx E_{\mathrm{Classic}}$, projected into both the Fock basis Eq. \eqref{HusFock} (b, e, h, k, n, q) and onto coherent states Eq. \eqref{Eq:Husimi} (c, f, i, l, o, r). In all cases, $U=0.7$ and $\epsilon=1.5$. In quantum projections, the number of particles varies: $N=130$ (b, k), $N=150$ (e, q), and $N=270$ (h, n).}
	 	\label{fig06b}
	\end{figure*}

Here, we draw a parallel between classical trajectories and Husimi functions for eigenvectors in the chaotic regime. 
In Figs \ref{fig06b}(a,d,g,j,m,p), the projections of six classical trajectories over the plane $(N_1,N_3)$ are shown, evolved for a short time at energy $E_{\mathrm{Classic}}$ for $\epsilon=1.5$, with six different initial conditions.
They are compared with the function $\mathfrak{H}_{E}(N_1,N_3)$ using Fock states in Figs.  \ref{fig06b}(b,e,h,k,n,q), and the projected Husimi function $\mathfrak{H}_{E}^{\mathcal{T}}(N_1,N_3)$ in Fig. \ref{fig06b}(c,f,i,l,o,r), for six different eigenstates with eigenvalues near the classical energy $E_{\mathrm{Classic}}$. Some more figures are in the appendix \ref{App1}.

Although the classical dynamics at energy $E_{\mathrm{Classic}}$ for $\epsilon=1.5$ is chaotic, it is possible to observe particular trajectories that are clearly recognized, where the density of points is larger, implying that the system spends more time in these regions of the parameter space. They could be associated with sticky orbits, like those found recently in chaotic billiards \cite{Lozej2021}. 

A careful revision of the projections of the functions $\mathfrak{H}_{E}(N_1,N_3)$ and $\mathfrak{H}_{E}^{\mathcal{T}}(N_1,N_3)$, belonging to hundreds of eigenstates with eigenenergies close to $E_{\mathrm{Classic}}$, 
allowed us to identify quantum interference patterns that closely resemble those observed in classical trajectories, under specific initial conditions. 
Similar correspondence between the projections of the Husimi functions of eigenstates have been reported for quasiperiodic orbits in the Dicke model of atoms in a cavity \cite{Pilatowsky2021,PVBLSH2021}. 
When the Husimi projections of eigenstates correspond to unstable periodic orbits, these projections are termed ``quantum scars." We are cautious about labeling our sticky orbits as unstable periodic orbits and our quantum projections as scars, 
especially since the highlighted classical orbits were selected among many other configurations. Furthermore, unstable periodic orbits are sets of null measure; other approaches are required to approximate them. We intend to study this question more thoroughly in the future.

The projection onto Fock states generally offers more detail in the distribution of probability than that offered by Husimi functions and requires less computational time. For this reason, only the projection into Fock states will be used for quantum analysis in what follows.
    
\subsection{Trajectory pattern}

\begin{figure}[h!]
    \centering
                \includegraphics[width=1\linewidth]{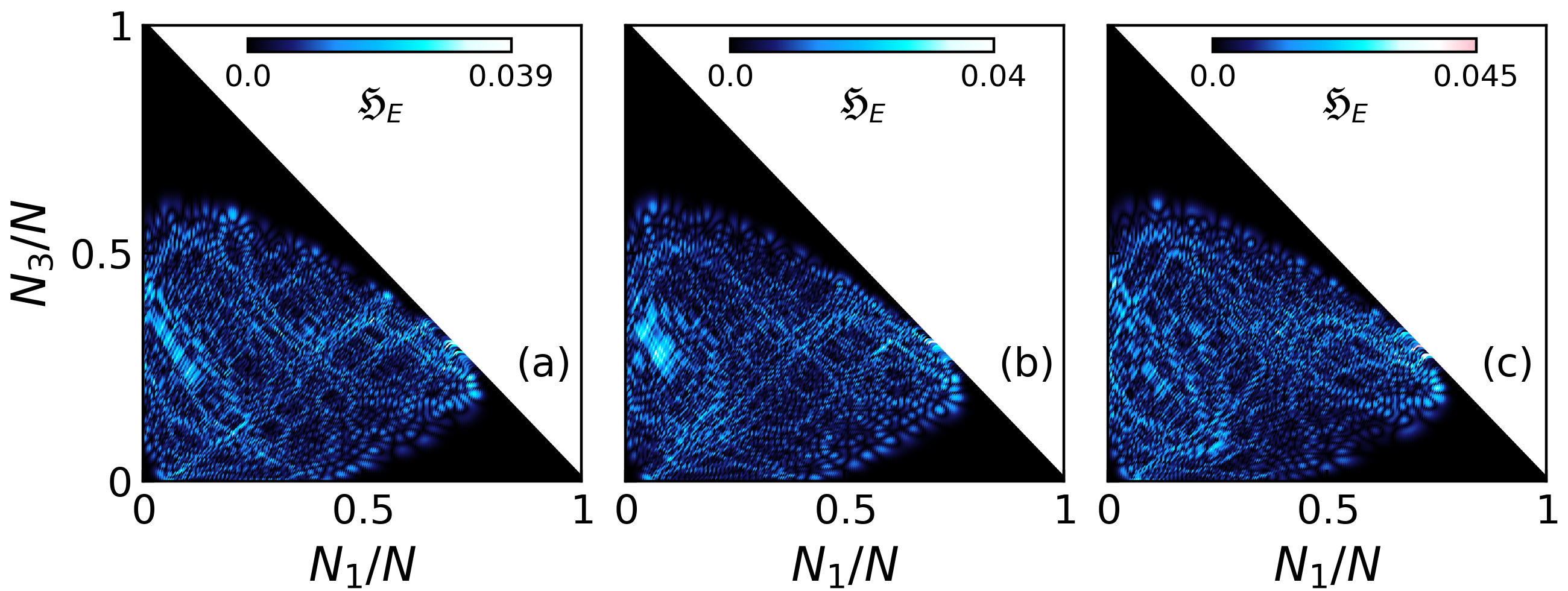}
        \caption{The function $\mathfrak{H}_{E}(N_1,N_3)$ \eqref{HusFock} of three eigenstates of a system with $N=210$, $U=0.7$, $\epsilon = 1.5$ and $E \approx E_{\mathrm{Classic }}$ that exhibit comparable projections.}
    \label{fig:claves1}
\end{figure}

\begin{figure}[h!]
    \centering
               \includegraphics[width=1\linewidth]{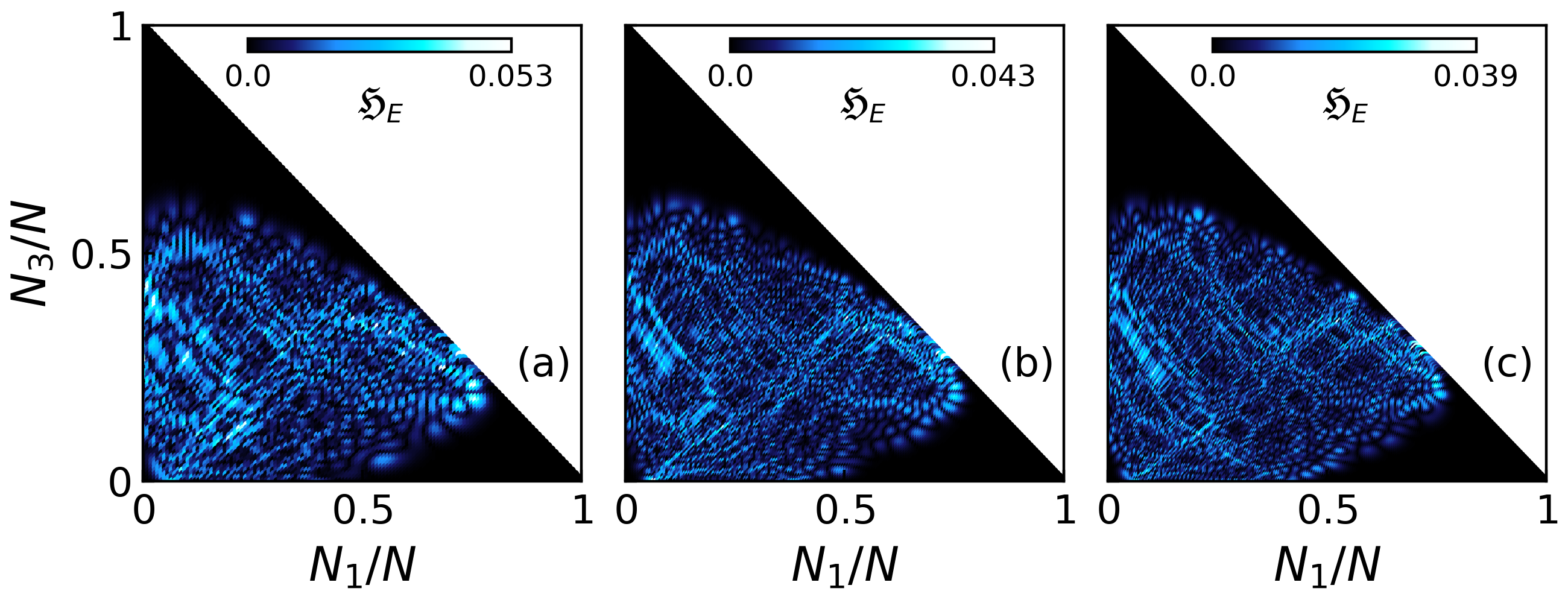}
        \caption{The function $\mathfrak{H}_{E}(N_1,N_3)$ \eqref{HusFock} for eigenstates of systems with varying numbers of particles: (a) $N=130$, (b) $N=180$, and (c) $N=210$, with energy $E \approx E_{\mathrm{Classic }}$. In all cases, $U=0.7$ and $\epsilon=1.5$.}
    \label{fig:claves2}
\end{figure}

\begin{figure}[htbp]
    \centering
       \includegraphics[width=1.\linewidth]{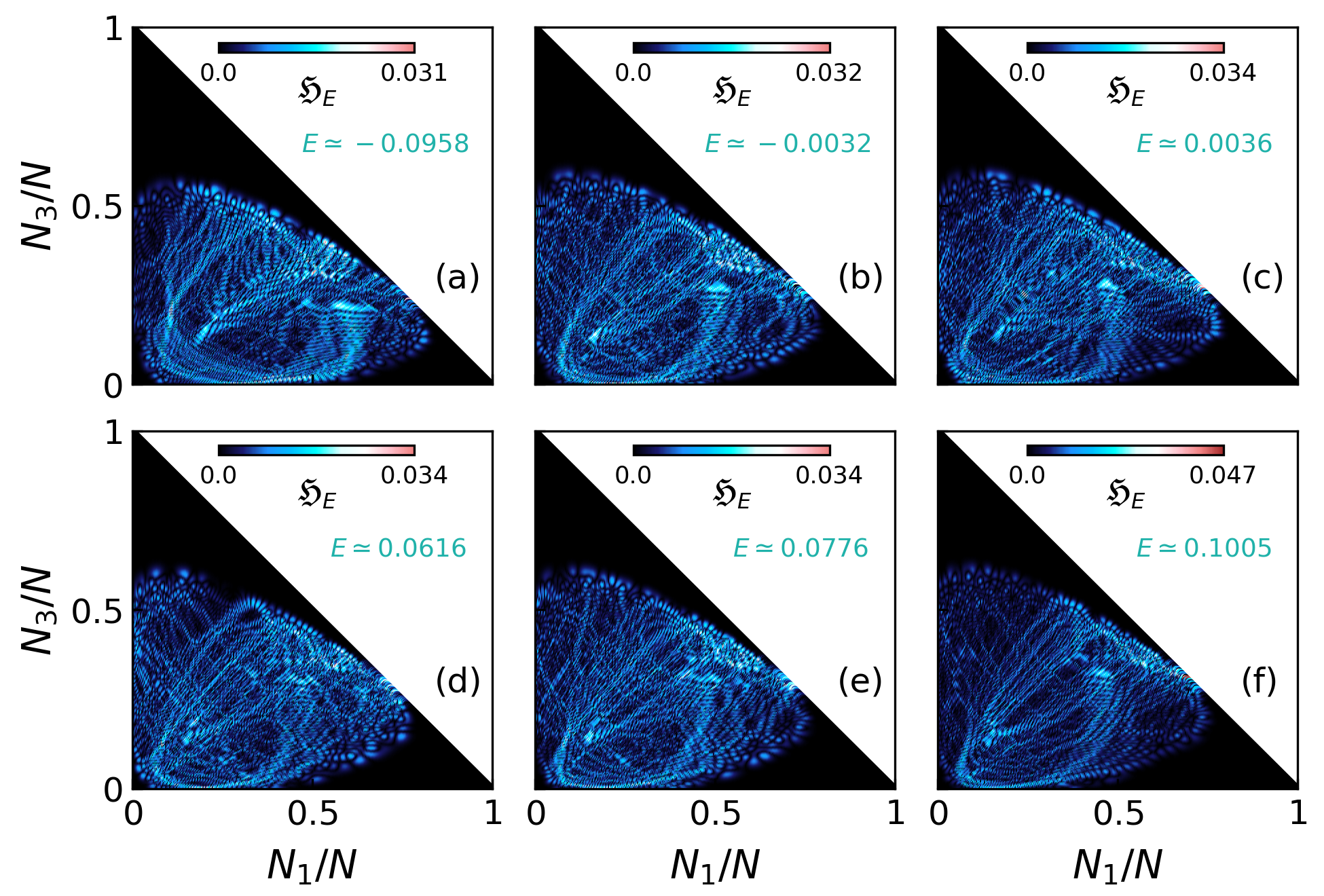}
        \caption{The function $\mathfrak{H}_{E}(N_1,N_3)$ for six specific eigenstates with eigenvalues (as indicated in the panels) distributed over a wider range of the energy spectrum, around $E_{\mathrm{class}}$.
        The distributions tend to be skewed, but it is possible to relate them to a standard shape. In all cases, $N=270$, $U=0.7$ and $\epsilon=1.5$. 
        }
    \label{fig:garras}
\end{figure}

The distinct shapes of the trajectories depicted in Fig. \ref{fig06b} were identified through visual similarity analysis, spanning the entire energy spectrum of systems with varying numbers of particles. These patterns emerge randomly for different eigenstates, even within the chaotic quantum regime, of the Hamiltonian \eqref{QH}. Each shape pattern can be found more than once within the chaotic regime of the system, as shown in Fig. \ref{fig:claves1} for $N=210$. They arise even when the number of particles in the system varies, as exemplified in Fig. \ref{fig:claves2}, for $N=130$, $N=180$, and $N=210$. It is also possible to identify the same pattern, sparsely, in a wider range of the energy spectrum, moving away from the critical chaotic energy $E_{Classci}$,
as exemplified in Fig. \ref{fig:garras}. Nonetheless, while it is possible to identify some other sets of related patterns, only a few emerge reasonably well-defined. Here, we present some of the most pronounced interference patterns (see Appendix \ref{App2} for some more examples).


The intriguing observation is that these quantum interference patterns appear recurrently also in classical trajectories for different initial conditions within the chaotic regime. This can be verified, for example, in Fig. \ref{fig00A}.
In this context, these projection patterns reflect a characteristic that pervades both classical and quantum systems within the bosonic model under investigation. This subject will be explored further in the next section.

We conclude this section with some additional observations: in Figs.(\ref{fig06b})-(\ref{fig:garras}), we observe that the interference patterns become better defined (continuous) as $N$ increases, in the case of projections onto Fock states, Eq. \eqref{Eq:Husimi}.
Interestingly, although the number of eigenstates per energy gap increases as $N$ increases, the search for eigenstates that unveil the relevant interference patterns is hampered. This is because the quantum correspondence of each classical trajectory is inherently tied to the energy of the respective eigenstate, and as $N$ increases, the system's sensitivity to energy variation also increases. Consequently, there is a need to seek states with increasingly precise energies. Paradoxically, the challenge of identifying such eigenstates seems to intensify with larger values of $N$. 
For this reason, the quantum projections with classical correspondence were primarily identified through several systems with different and relatively small numbers of particles.

\section{The classical-quantum correspondence}\label{section2}
This section explores the quantum-classical correspondence for the 3-well model in the regular ($\epsilon=0$), mixed ($\epsilon=0.7$), and chaotic ($\epsilon=1.5$) regimes, with $U=0.7$ and $J=1$. The analysis will center around the energy $E_{\mathrm{Classic}} \approx 0.0752$ (the critical point energy studied in ref. \cite{CCRSH2021}), as in the preceding sections.
	
	
	For the classical dynamics, the plots $N_1/N$ vs $\phi_{12}$ on the plane $\phi_{32} = 0$ are Poincar\'e sections, useful to visualize the regular, mixed or chaotic trajectories. These sections are constructed in two ways: using hundreds of initial conditions evolved for short times and a few initial conditions evolved for long times. They provide similar yet complementary views of classical dynamics. Some of these initial conditions are evolved in time to obtain projected trajectories in coordinates $N_3/N$ vs $N_1/N$, which are then compared with the quantum probability distributions $\mathfrak{H}_E(N_1,N_3)$.	
	
	\subsection{The integrable case}
	
	We select the parameter values $\epsilon=0$, $U=0.7$,
        and $J=1$ to study the integrable case. 
	
	\subsubsection{The conserved quantity $\hat{Q}$}
	
	The quantum operator $\hat{Q}$ %
	\begin{equation}
		\hat{Q}=\left(\hat{N}_1+\hat{N}_3\right)-\left(a_1^{\dagger}a_3+a_3^{\dagger}a_1\right)
	\end{equation}
    is in this case a constant of motion, satisfying $\left[\hat{Q},\hat{H}\right]=0$. $\hat{Q}$ can be obtained directly using the Bethe ansatz~\cite{WYTLA2018}.
    There are $N+1$ eigenvalues $q_m=2(m-1)/N$, $m=1,...,N+1$ of $\hat{Q}/N$, equally distributed in the interval $[0,2]$, through $q_1=0$ and $q_{N+1}=2$.
    Each eigenstate $|E\rangle$ is also an eigenvector of $\hat{Q}/N$. Different eigenstates have the same eigenvalues $q$ of $\hat{Q}/N$, with a degeneracy $m$-dependent: $N+2-m$. 
	
		The classical limit of $\hat{Q}$ in normalized coordinates $\rho_k=\sqrt{N_k/N}$ is
	
	\begin{equation}
		Q=\rho_1^2+\rho_3^2-2\rho_1\rho_3\cos\left(\phi_{32}-\phi_{12}\right)
	\end{equation}
	$Q$ takes all the possible values in the interval $\left[0,2\right]$. Particularly, $Q=0$ when $N_2/N=1$ and $Q=2$ when $N_1/N=N_3/N=1/2$, $\phi_{32}-\phi_{12}= (2n+1)\pi$, with $n \in \mathbb{Z}$.
	
	Two classical conserved quantities exist: the total energy and the classical $Q$, implying two natural frequencies. This work focuses on quantum-classical correspondence, so we do not calculate the frequencies. However, the evolution of many initial conditions suggests that the frequencies are incommensurable and manifest in a quasi-periodic motion.

The quantum-classical correspondence is straightforward in this case: any eigenstate with eigenvalue $q$ of $\hat{Q}/N$ has a function $\mathfrak{H}_{E}(N_1, N_3)$ that is essentially identical to a projected classical trajectory with the classical value $Q=q$ and a classical energy value very close to $E$.
	
	\subsubsection{Classical trajectories}
	
\begin{figure}
		\centering
   		\includegraphics[width=1.\linewidth]{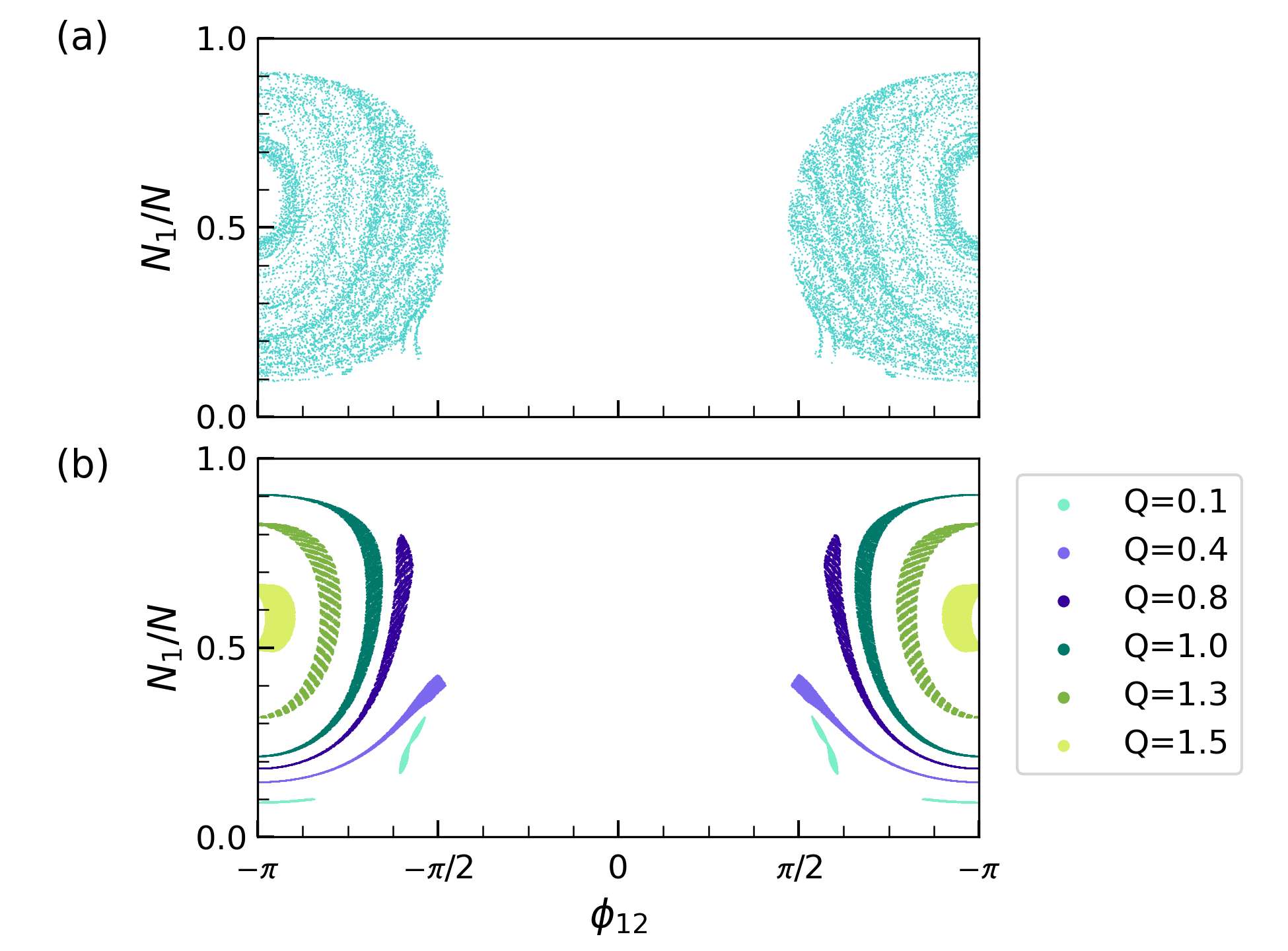}
     \caption{
     	Poincar\'e section with energy $E=E_{\mathrm{Classic}}$ for the coordinates $(N_1,\phi_{12})$, with $\phi_{32}=0$; (a) for 460 trajectories, each evolved in short time intervals ($t_f=100$), (b) for 6 trajectories with different values of $Q=0.1, 0.4, 0.8, 1.0, 1.3$, and $1.5$ respectively, evolved in large time intervals $t_f=10000$. In both cases, $U=0.7$ and $\epsilon=0$ (integrable regime).}
   \label{fig02}
	\end{figure}

 The integrability and quasiperiodicity allow for the classification of all classical trajectories. For each classical value of $Q$, there is a unique trajectory encompassing a specific region. Figure \ref{fig02} b) displays the intersection of six trajectories indexed by six different values of $Q$ and the hypersurface $\phi_{32}=0$. The points are projected onto the coordinates $(N_1, \phi_{12})$. Well-localized bands are observed, associated with quasiperiodic orbits.
    
   Fig.\ref{fig02} a) is a similar Poincar\'e section, for many trajectories with initial conditions distributed in all the hypersurface evolved for short times. The patterns are similar, but the bands are harder to observe.
	
	To explore the quantum-classical correspondence, the classical trajectories at long times projected in the coordinates $(N_1, N_3)$ are compared with the quantum projection $\mathfrak{H}_E(N_1, N_3)$. 
	
		\begin{figure}
		\centering
                 \includegraphics[width=0.95\linewidth]{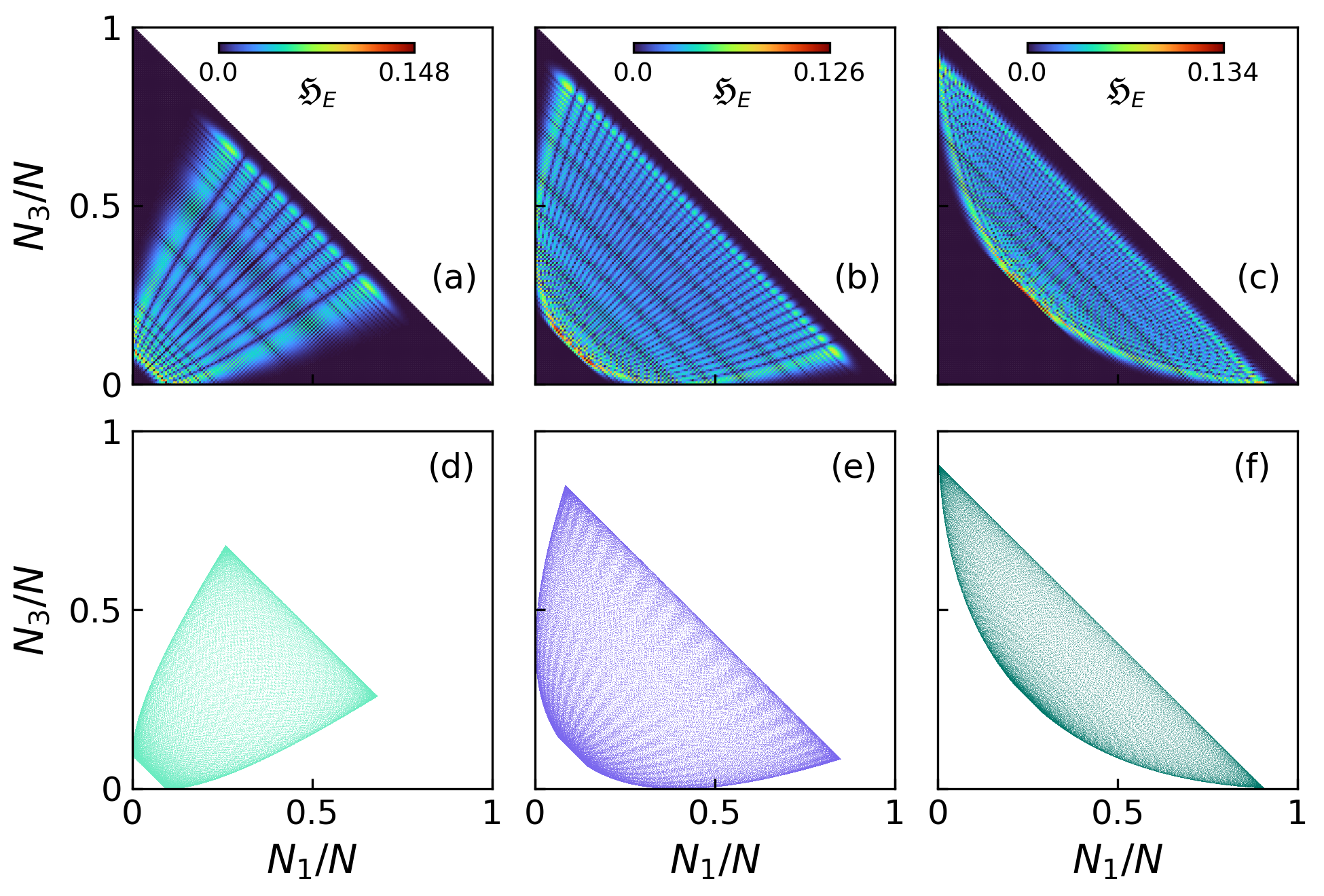}
            \caption{Comparison between quantum projections (upper) and classical trajectories (bottom) for the integrable regime ($\epsilon=0$). (a, b, c): Function $\mathfrak{H}_E(N_1, N_3)$ for three eigenstates of $\hat{Q}/N$, with eigenvalues $q=0.1, 0.4, 1$, respectively. (d, e, f): Classical trajectories projected onto the plane $(N_1, N_3)$ for three initial conditions, with $Q=0.1,0.4$ and $1.0$, respectively. (see also these three cases in Fig. \ref{fig02}).}
		\label{fig03}
	\end{figure}

Figure \ref{fig03} is an example of the quantum-classical correspondence for $\epsilon=0$. The function $\mathfrak{H}_E(N_1,N_3)$ for three eigenstates with energies closest to $E_{\mathrm{Classic}}$ and three different values of $q$ (respectively 0.1, 0.4, and 1, from left to right) is compared with three classical trajectories with the same value of $Q$ and classical energy $E_{\mathrm{Classic}}$ projected onto the plane $(N_1, N_3)$. The remarkable similarity between both descriptions and the fringes resembling an interference pattern are also visible in the classical plots.
	
        \subsection{The mixed case $\epsilon=0.7$}
	
	In this section, we study the case $\epsilon=0.7$, which shows a mixed behavior. 
	
\begin{figure}
		\centering
    \includegraphics[width=1.\linewidth]{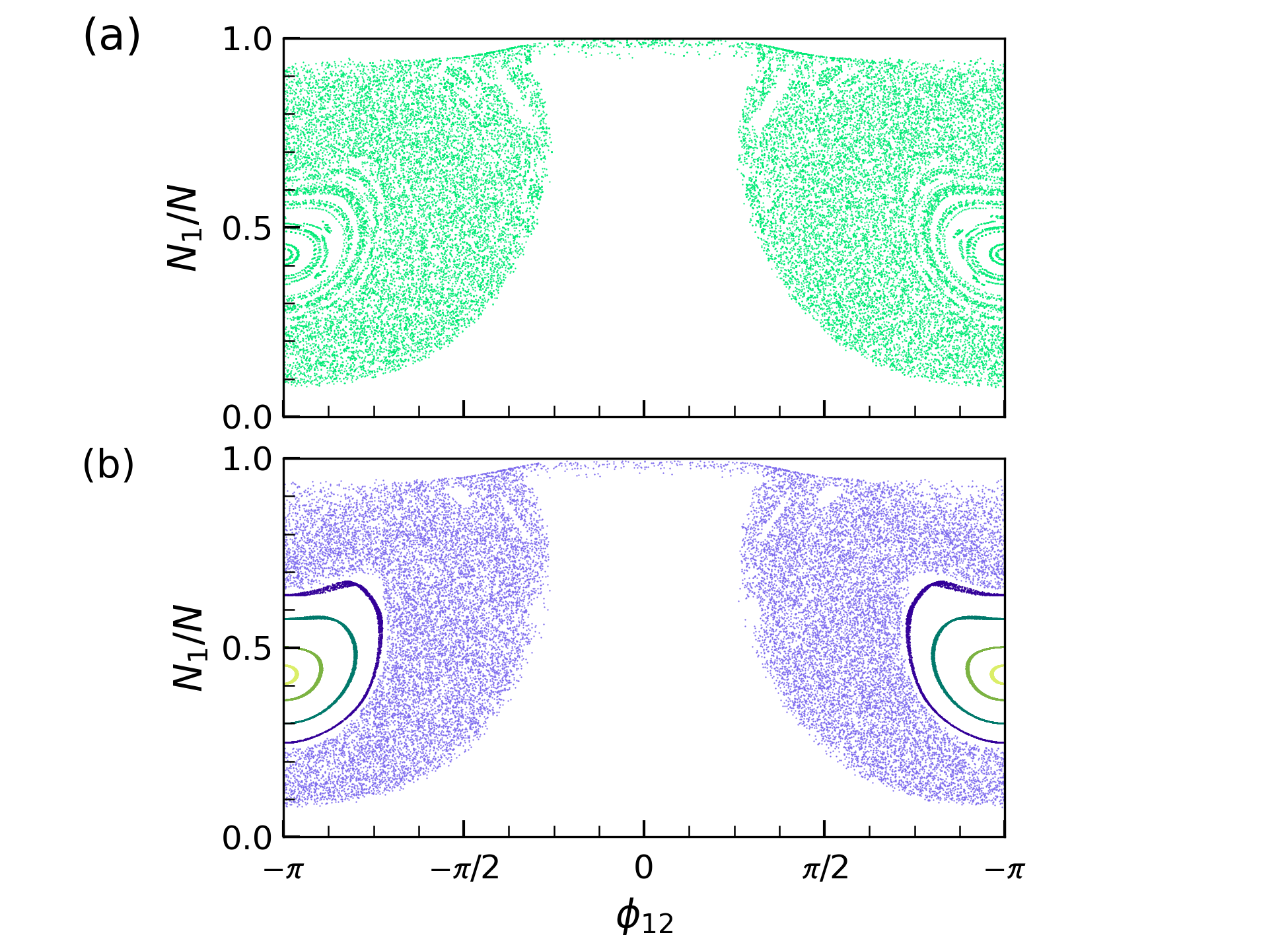}
		\caption{Poincar\'e sections for $\epsilon=0.7$ (mixed regime), $E=E_{\mathrm{Classic}}$ and $\phi_{32}=0$. a)  $513$ trajectories homogeneously distributed and evolved for short times. b) Five initial conditions evolved for long times.}
		\label{fig04}
	\end{figure}

	The coexistence of regular and chaotic regions can be observed in the Poincar\'e sections shown in Fig.  \ref{fig04}, displaying the results for the fixed energy $E_{\mathrm{Classic}}$ and $\phi_{32}=0$. Fig.  \ref{fig04} (a) was made using 513 initial conditions evolved at short times, and Fig.  \ref{fig04} (b) for five initial conditions at long times. The regions of regular dynamics with quasi-periodic trajectories and the chaotic ones are recognized in both figures. The purple points set up a typical chaotic region, and the other four trajectories configure well-localized tiny strips.

\begin{figure}
		\centering
    \includegraphics[width=1.\linewidth]{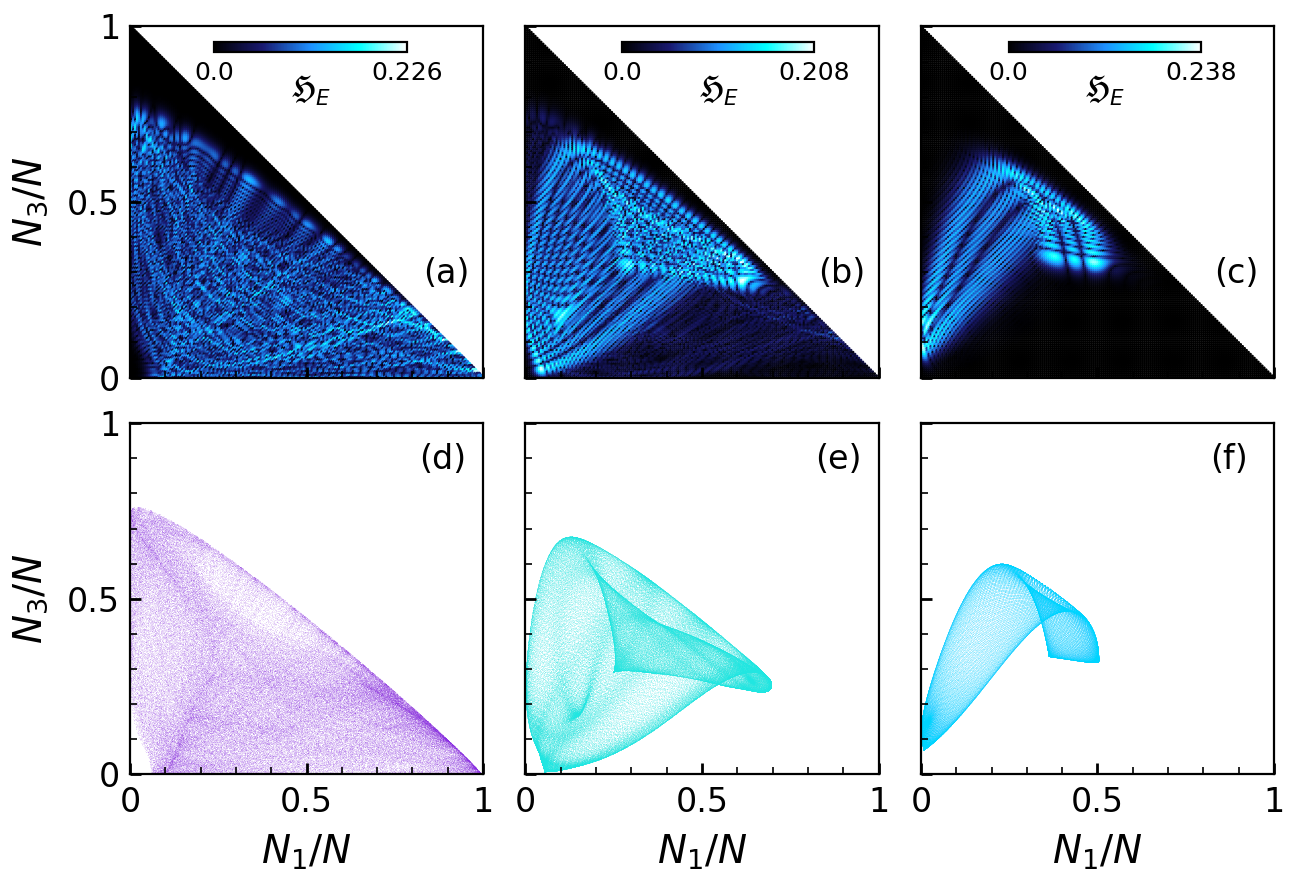}
		\caption{Upper panels (a, b, c):  Function $\mathfrak{H}_E(N_1,N_3)$ for three eigenstates with eigenvalues closest to $E_{\mathrm{Classic}}$. Bottom panels (d, e, f): Projected classical trajectories onto the plane $(N_1, N_3)$ for initial conditions with $E=E_{\mathrm{Classic}}$ and $\epsilon=0.7$.}
		\label{fig05}
	\end{figure}

	For $\epsilon=0.7$, analyzing the quantum-classical correspondence is more challenging. On the classical side, the five trajectories in Fig. \ref{fig04} (for all times) are projected onto the coordinates $(N_1, N_3)$. Three are presented at the bottom of Fig. \ref{fig05}. The trajectories are projected onto the coordinates $(N_1, N_3)$ for times from $t=0$ to $t=10000$. The purple classical trajectory is chaotic, and the other two are regular and quasiperiodic. For the quantum analysis, 200 eigenstates closest to $E_{\mathrm{Classic}}$ calculated with $N=180$ were selected, and their functions $\mathfrak{H}_E(N_1, N_3)$ were calculated. Most of the 200 eigenstates extend in the same region as the purple trajectory; only one is presented. A few eigenstates in this energy region have a localized function $\mathfrak{H}_E(N_1, N_3)$, resembling the classical projections. Two of them are shown in Fig. \ref{fig05}(b) and \ref{fig05}(c).
	
	\subsection{The chaotic case $\epsilon=1.5$}

\begin{figure}
		\centering
        	 	\includegraphics[width=0.9\linewidth]{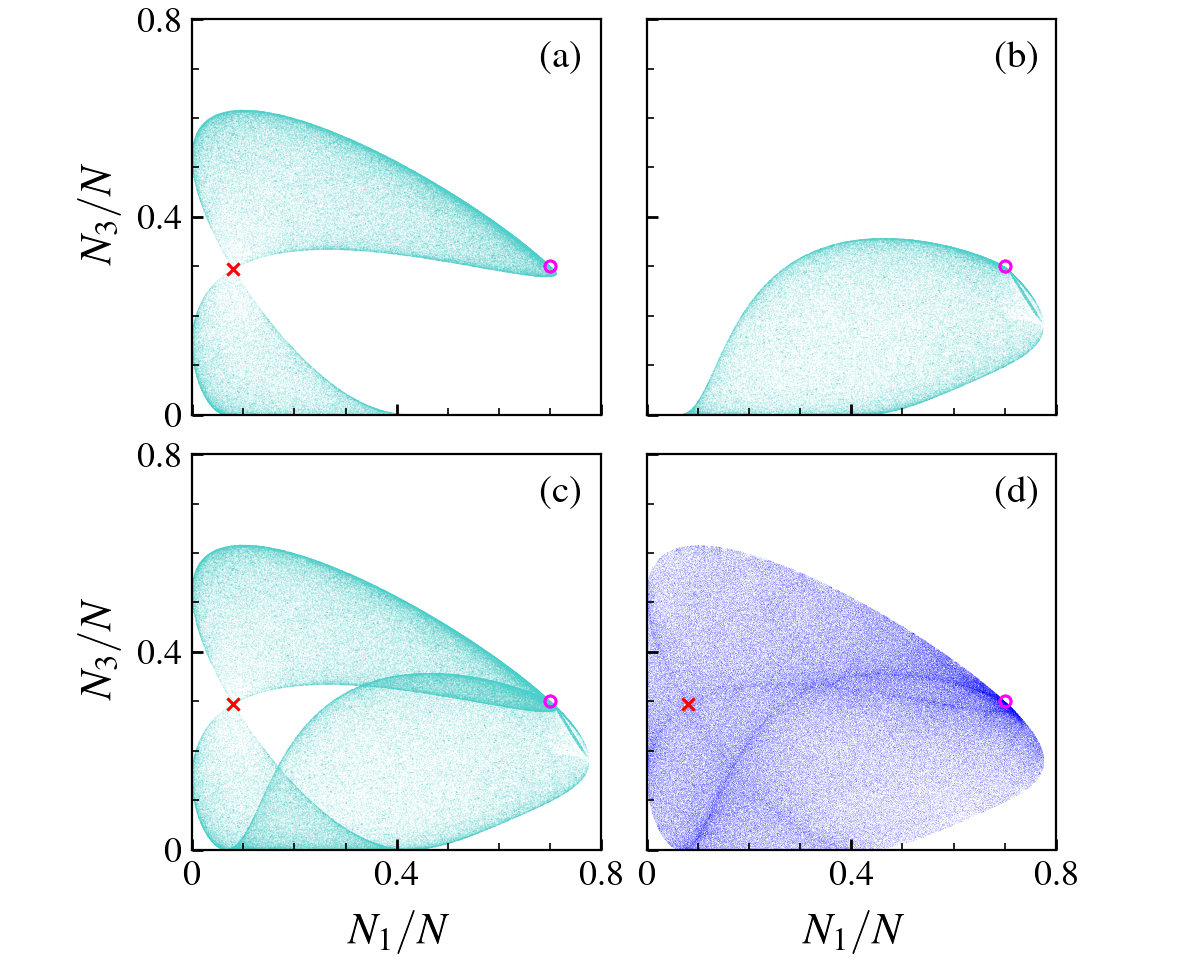}
           		\caption{Poincar\'e sections in coordinates $(N_1/N, N_3/N)$ for $(E,\phi_{32})=(E_{\mathrm{Classic}},\pi)$ (a) and $(E,\phi_{32})=(E_{\mathrm{Classic}},0)$ (b). An overlap of both (a) and (b) Poincar\'e sections is shown in (c). Panel (d) depicts an arbitrary classical
                chaotic trajectory projected onto $(N_1, N_3)$ coordinates for long times. Note that the pattern of the Poincar\'e
                section in panel (c) is reproduced in the projection. The 'x' points mark the classical unstable critical point, while the magenta circles mark the prevailing initial state of the classical trajectories. In the figure, $\epsilon=1.5$ (chaotic regime).}
		\label{fig06}
	\end{figure}

The study of the case with $(U, J, \epsilon) = (0.7, 1, 1.5)$ reveals explicit quantum chaos behavior in the middle of the energy spectrum region (particularly at $E/N \approx 0.0752$), and in this case, the associated classical scenario exhibits a typical chaotic Poincaré section (see Fig. \ref{fig06}). For any initial condition with energy $E = E_{\mathrm{Classic}} \approx 0.0752$, each trajectory encompasses all accessible phase space and has the same long-time structure.

	Fig. \ref{fig06}(a) displays the Poincaré section on $(N_1/N, N_3/N)$ with energy $E_{\mathrm{Classic}}$, but restricted to $\phi_{32}=\pi$, obtained from one initial condition at long times. It covers two areas connected at one point, which coincides with the unstable equilibrium points studied in \cite{CCRSH2021}. Fig. \ref{fig06}(b) displays the Poincaré section restricted to $\phi_{32}=0$. Both exhibit a noticeable concentration of points in their external contour, where the values of $\phi_{12}$ are also very close to ${0,\pm\pi}$.
	 
	As shown in ref. \cite{CCRSH2021}, any classical critical point has exactly these extremal phase values. As a consequence, any chaotic trajectory spends more time close to these contour points since, in them, $\partial \mathcal{H}/\partial \phi_{i2}=0$ with $i\in {1,3}$. Fig. \ref{fig06}d) displays the $(N_1, N_3)$ projection of an evolved trajectory from an arbitrary initial condition. The density of points is higher in the vicinity of the critical regions.
	  
	To analyze the quantum description of the phase space in the chaotic regime,  it is useful to calculate the average of the distributions $\mathfrak{H}_{E}(N_1,N_3)$ in an energy window 
    $\mathfrak{E} $  in the interval $\left(E_{\mathrm{Classic}}-\Delta E/2,E_{\mathrm{Classic}}+\Delta E/2\right)$:
	 \begin{equation}\label{Prom}
	 \bar{\mathfrak{H}}(N_1,N_3)=\frac{1}{\mathcal{N}}\sum_{E \in \mathfrak{E}}\mathfrak{H}_{E}(N_1,N_3)
	 \end{equation}

\begin{figure}
	 	\centering
           	 	\includegraphics[width=1\linewidth]{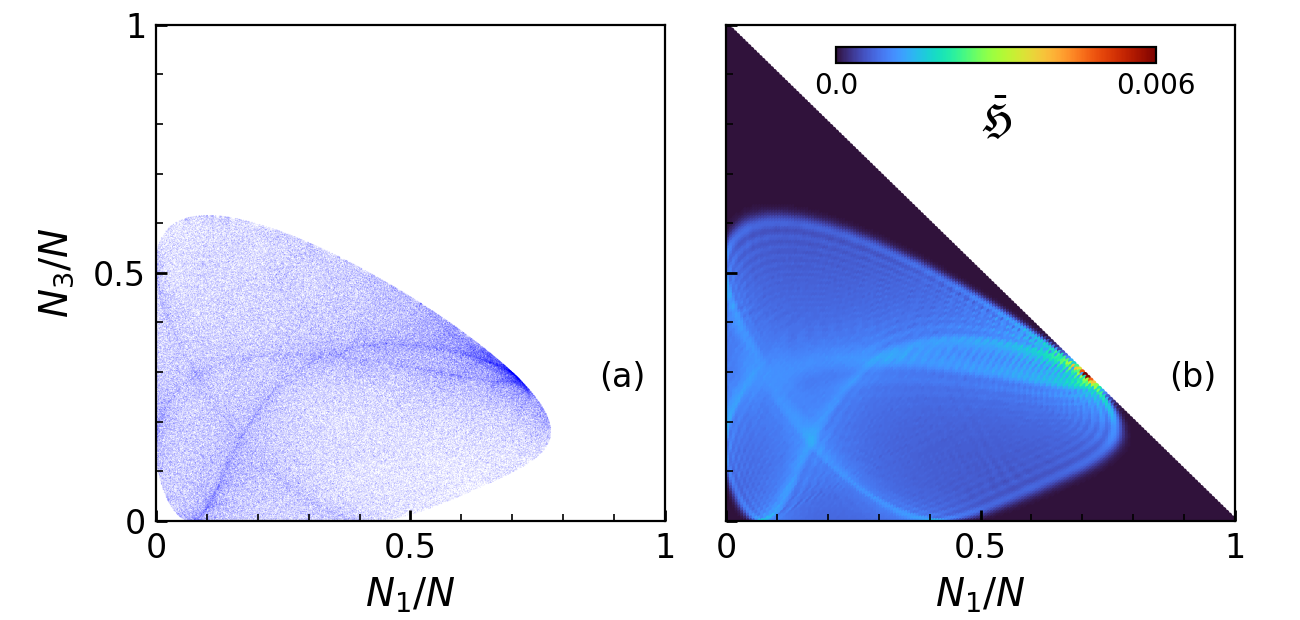}
       \caption{Panel (a) displays an arbitrary classical chaotic trajectory with an energy of $E/N\approx 0.0752$. In panel (b), the mean value distribution, represented by Eq. \eqref{Prom}, is presented. This distribution encompasses all eigenstates falling within an energy window of $\Delta E=0.02$ (comprising 172 eigenstates) centered at $E_{\mathrm{Classic}} \approx 0.0752$ for a system of size $N=180$. It is noteworthy to observe the remarkable correspondence between the density distributions of the classical and quantum projections. In both panels, $U=0.7$ and $\epsilon=1.5$ (chaotic regime).}
       \label{fig09}
	 \end{figure}

It is shown for $N=180$, with $\mathfrak{E} $ containing 172 eigenstates around $E/N \approx 0.0752$ with $\Delta E=0.02$ in Fig. \ref{fig09}.

  It is remarkable that the average over a small energy window of the projection of the quantum eigenstates closely reproduces the phase space structure observed for any chaotic trajectory when projected onto the coordinates $(N_1, N_3)$. In this chaotic regime, the long-time behavior of any classical trajectory with classical energy $E_{\mathrm{Classic}}$ is reflected in the microcanonical average of the projected Husimi function around the nearest $\mathcal{N}$'s eigenenergies to $E_{\mathrm{Classic}}$. Any classical chaotic trajectory spends more time in the dense region shown in Fig. \ref{fig09}, which is directly connected with the unstable critical point with energy $E/N \approx 0.0752$ having the same values of $(\phi_{12},\phi_{32})=(0,\pm \pi)$.

   \begin{figure}
	 	\centering
         \includegraphics[width=1\linewidth]{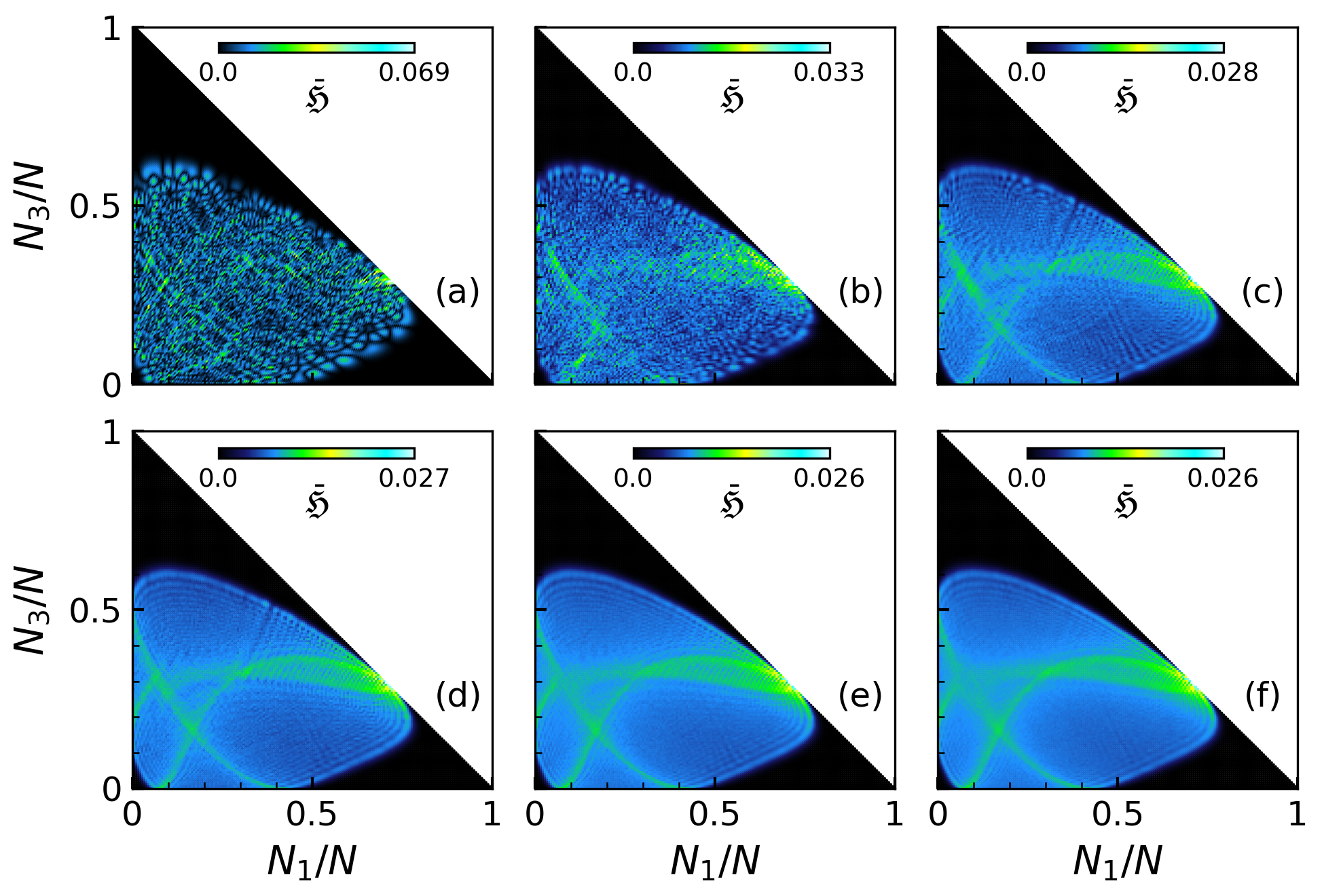}
	 	\caption{The mean value distribution, given by Eq. \eqref{Prom}, is showcased across various energy windows, each centered at $E_{\mathrm{Classic}} \approx 0.0752$. These windows correspond to different numbers of eigenstates: (a) 1, (b) 10, (c) 50, (d) 100, (e) 200, and (f) 300, for a system of size $N=180$, $U=0.7$ and $\epsilon=1.5$.}
	 	\label{figMean}
	 \end{figure}

 In Fig. \ref{figMean}, we display the mean value (\ref{Prom}) for different values of $\Delta E$ with $N=180$ and eigenvalues $E/N$ closest to $E_{\mathrm{Classic}}$. Each energy interval contains a different number of associated eigenvectors. It is clear from the figure that the characteristic pattern in the average (\ref{Prom}) is generated for different $\mathcal{N}$, even present when $\mathfrak{E}$ contains only ten eigenvectors (Fig. \ref{figMean}b). Fig. \ref{figMean}e shows a more fine contrast, having $\mathcal{N}=200$. On the other hand, for $\mathcal{N}=300$, the pattern begins to defocus, mainly in the region around the critical point.

\begin{figure}
	 	\centering
    \includegraphics[width=1.\linewidth]{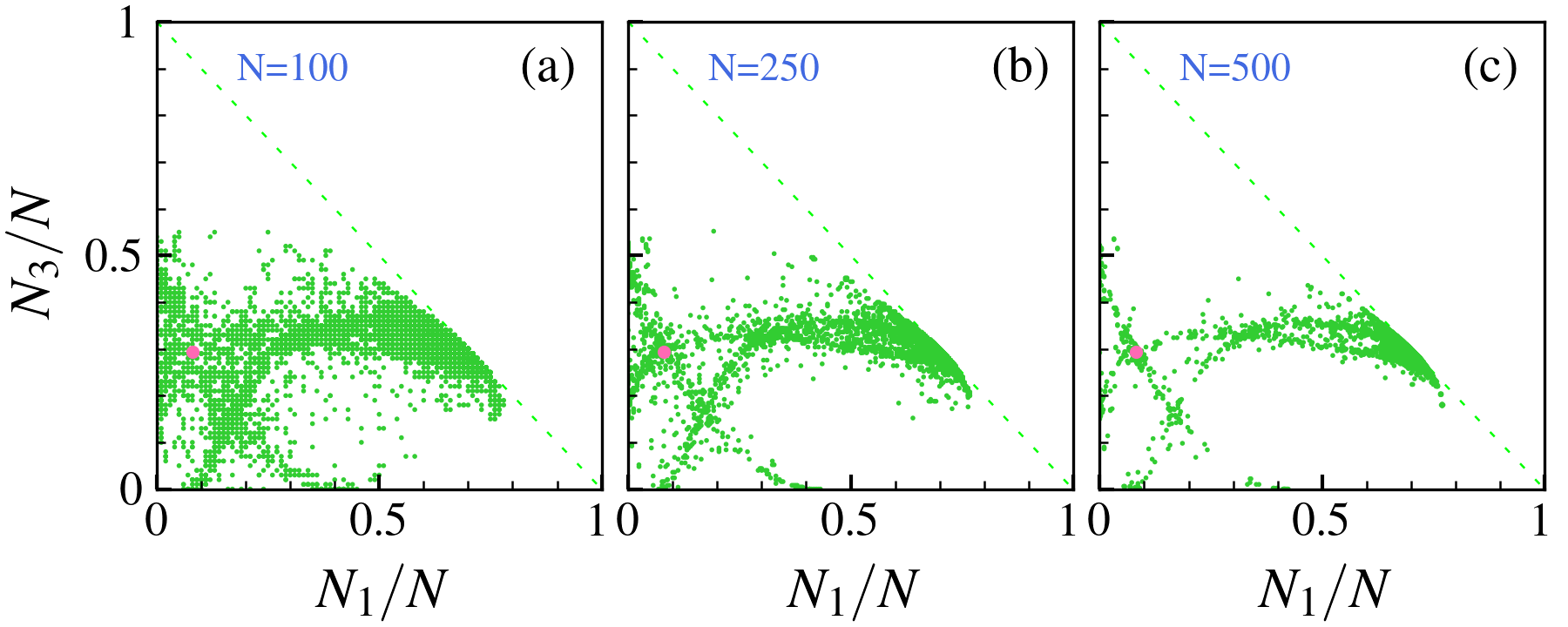}
	 \caption{
  $(N_1/N, N_3/N)$ indices of the Fock states with the most prominent components in the 200 eigenstates used in Eq. \eqref{Prom}. The total number of particles in each case is (a) $N=100$, (b) $N=250$, and (c) $N=500$. From each eigenstate, 20 components were selected. It can be seen that the classical and quantum dense set region in Fig. \ref{fig09} is reproduced. In all cases, $U=0.7$ and $\epsilon=1.5$.}
	 	\label{PrincipalFocks}
	 \end{figure}

   Maximally chaotic eigenstates exhibit a specific Gaussian distribution of their components; therefore, mean values like (\ref{Prom}) would eliminate any specific pattern of the particular eigenstate distributions. The more chaotic eigenstates do not follow a perfectly Gaussian distribution of the components, although the system satisfies all the spectrum criteria of chaotic quantum systems \cite{WCFS2022}. Particularly, the deviation from Gaussian behavior is more evident at the extremum of the distribution, where the more prominent components reside. This can be interpreted as a remnant localization, which is not apparent for individual eigenstates but is evident in the mean value (\ref{Prom}) where many close eigenvalues are present. The dense region in figure \ref{fig09}b expresses the slight deviation from the total delocalization characteristic of chaotic eigenstates. The Fock states in the dense region contribute more to the eigenstate superposition. This is visible in Fig. \ref{PrincipalFocks}, where we show the 20 Fock states with the largest absolute value contribution in each one of the $\mathcal{N}=200$ eigenstates used in (\ref{Prom}). The 4000 Fock states, with the principal contribution, constitute the dense region of Fig. \ref{fig09}b.

\section{Discussion}\label{section5}

In this work, we have established the quantum-classical correspondence of a many-body bosonic system with an integrable limit. The model is sufficiently simple to apply many techniques of the quantum-chaos discipline but has some particularities, such as the quasiperiodic behavior of the classical trajectories and the non-standard distribution of the chaotic states. The transition to chaos is demonstrated, and a detailed comparison between classical and quantum is carried out. In the integrable limit, a complete classification of the orbits is recognized, and the associated projected Husimi function of an eigenstate with energy very close to the classical energy is calculated. In particular, we find an eigenstate with energy very close to $E_{\mathrm{Classic}}$ for sufficiently large $N$ associated with any classical quasiperiodic trajectory. The transition to chaos mixes the quasiperiodic orbits and the associated quantum states in a complicated way; the quantum states are now related to intricate trajectories only in short time intervals. Remarkably, the large-time structure of any chaotic trajectory is reproduced by the averaged Husimi function, which represents a dense region connected with the unstable critical point that characterizes this larger time structure. The dense region is also related to the Fock states with the larger contribution in each one of the eigenstates belonging to $\mathfrak{E}$, used in the Husimi mean value (\ref{Prom}). Our study demonstrates how a slight deviation from perfectly delocalized eigenstates influences the large-time behavior of classical chaotic trajectories and establishes the quantum-classical correspondence firmly, even in the chaotic case.

\section{acknowledgment}

E. Castro thanks the Brazilian CNPq agency for partial financial support. AF and KW acknowledge support from CNPq (Conselho Nacional de Desenvolvimento Científico e Tecnológico) - Edital Universal 406563/2021-7. IR also thanks CNPq for partial support through contract 311876/2021-8. J.G.H acknowledge partial financial support from project PAPIIT-UNAM IN109523.
We use the Julia free software TaylorIntegration.jl for the dynamics of all the classical trajectories. We thank Lea Ferreira dos Santos for valuable discussions and suggestions.


\appendix

\section{Quantum projection patterns: Additional figures}\label{App2}



\begin{figure}[h!]
    \centering
                \includegraphics[width=1.\linewidth]{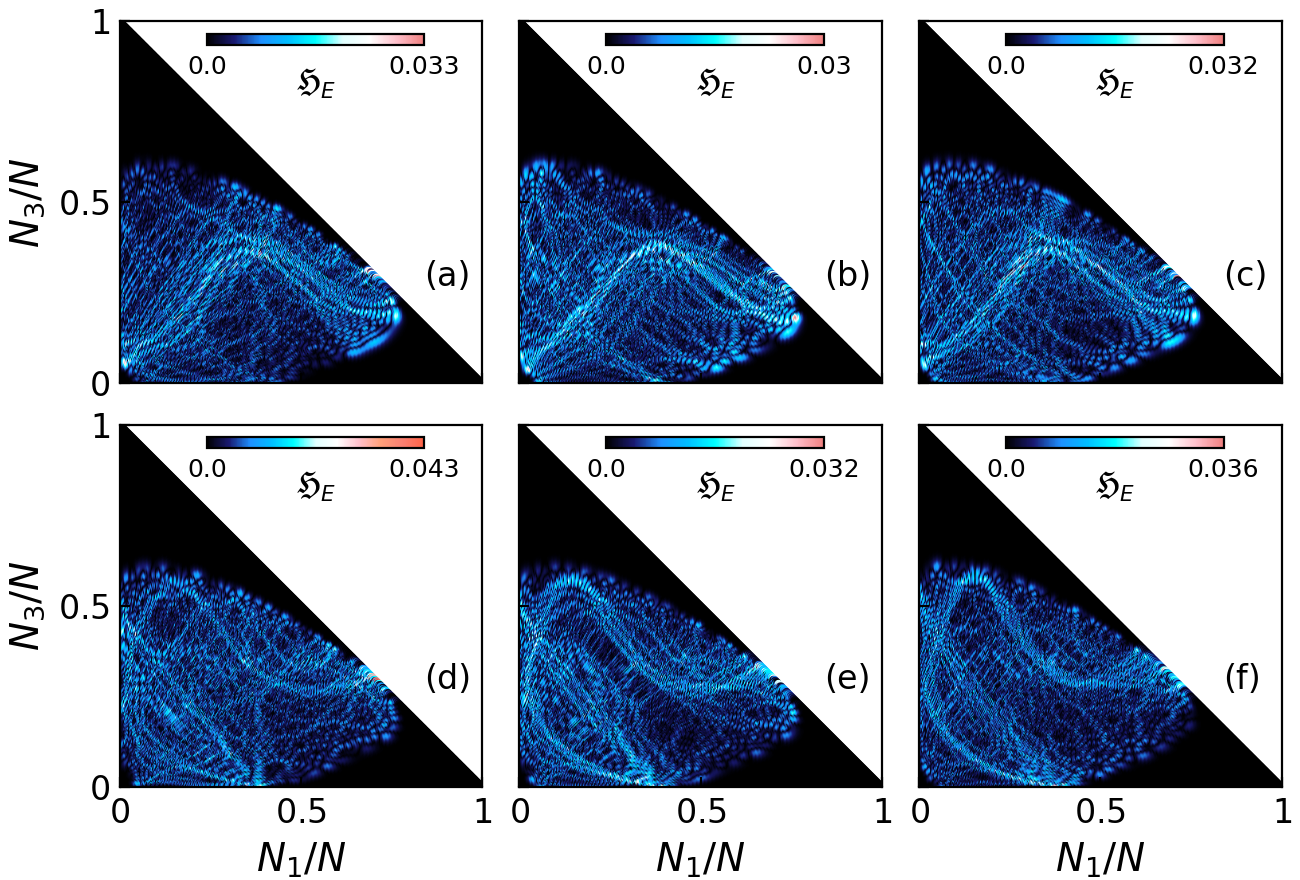}
        \caption{The function $\mathfrak{H}_{E}(N_1,N_3)$ of two particular sets of eigenstates that exhibit similar distributions, with $E \approx E_{\mathrm{Classic}}$, for $N=270$, $U=0.7$ and $\epsilon=1.5$.}
    \label{fig:a21}
\end{figure}
\begin{figure}[h!]
    \centering
                     \includegraphics[width=1.\linewidth]{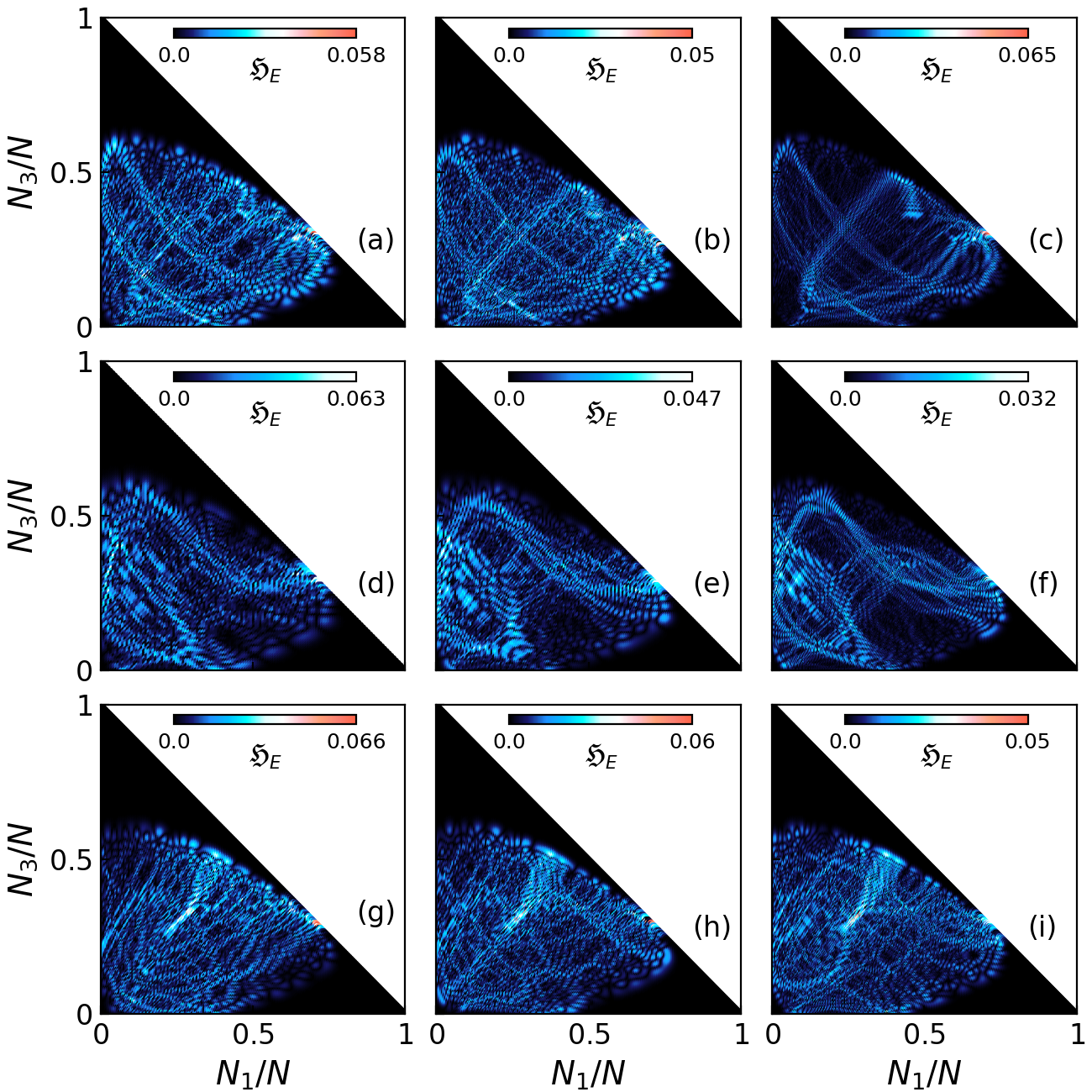}
        \caption{The function $\mathfrak{H}_{E}(N_1,N_3)$ of a set of eigenstates, with $E \approx E_{\mathrm{Classic}}$, that display similar distributions for systems with different numbers of particles, respectively: (upper) $N=180$, $N=220$ and $N=270$. (middle) $N=130$, $N=160$ and $N=260$. (bottom) $N=180$, $N=200$ and $N=240$. For all, $U=0.7$ and $\epsilon=1.5$.}
    \label{fig:a22}
\end{figure}

Here, we present additional examples of distinct forms that prominently emerge in the projections of the functions, $\mathfrak{H}_E\left(N_1,N_3\right)$, for quantum chaotic systems. These projections correspond to eigenvectors with energies close to the value of the unstable classical critical point, $E \approx E_{\mathrm{Classic}}$.
In Fig. \ref{fig:a21}, two recurring shape patterns are highlighted, emerging through different eigenvectors from the most chaotic region of the quantum system spectrum, with a fixed $N=270$. Conversely, in Fig. \ref{fig:a22}, shape patterns resulting from eigenvectors of systems with different numbers of particles are presented.

\section{Classical-quantum correspondence patterns: Additional figures}\label{App1}

\begin{figure}[h!]
	\centering
                 \includegraphics[width=1.\linewidth]{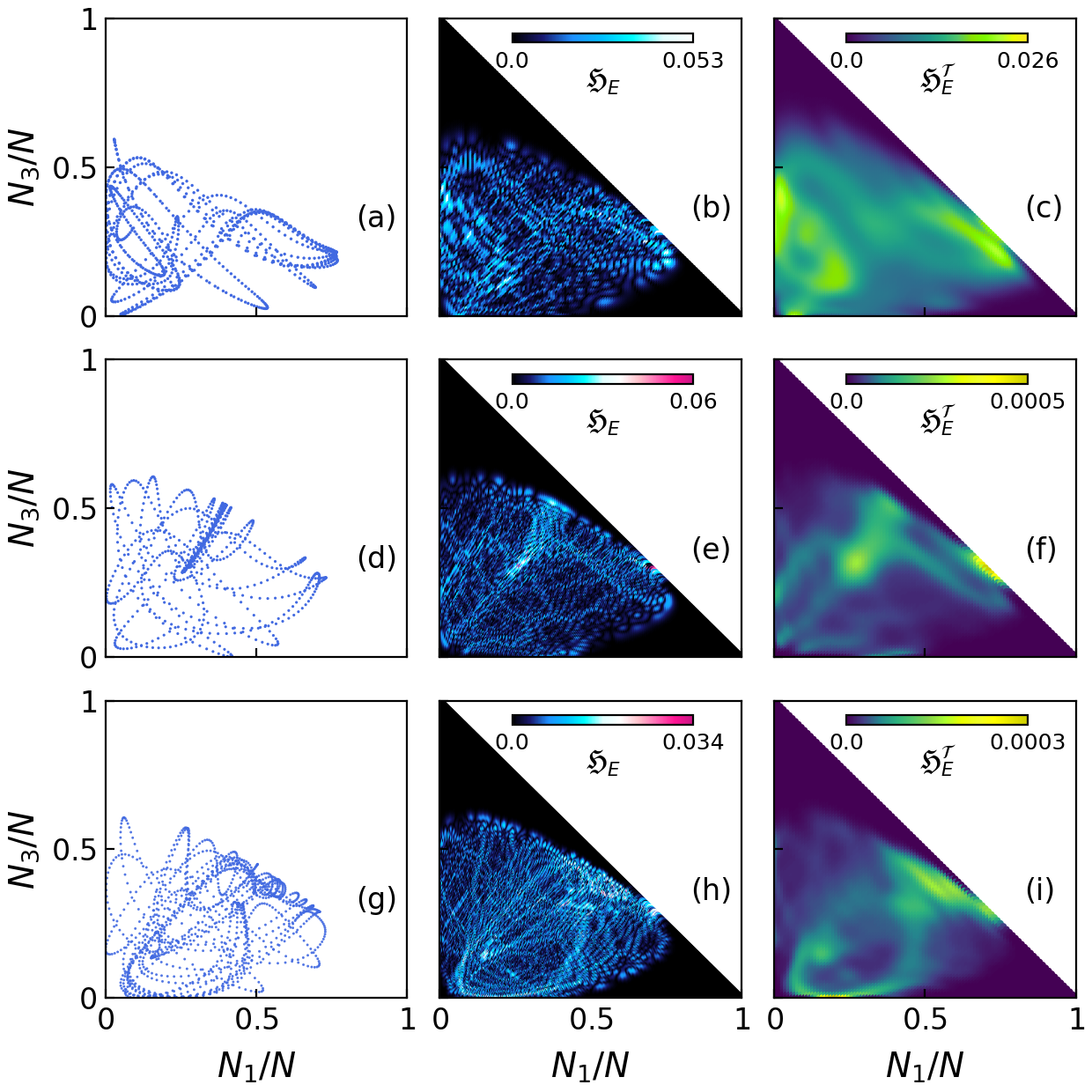}
	\caption{Projections of three classical trajectories onto the plane $(N_1,N_3)$ for different initial conditions with energy $E_{\mathrm{Classic}}$ (a, d, g). Equivalent projections are obtained for the function $\mathfrak{H}_{E}(N_1,N_3)$ (b, e, h) and $\mathfrak{H}_{E}^{\mathcal{T}}( N_1,N_3)$ (c, f , i), for eigenstates with energy $E\approx E_{\mathrm{Classic}}$. In quantum projections, the number of particles varies: $N=130$ (top row), $N=200$ (middle row), and $N=270$ (bottom row). In all cases, $U=0.7$, $\epsilon=1.5$, and $J=1$.}
	\label{fig:appB}
\end{figure}

It is possible to identify classical-quantum correspondence for all the standard forms shown in this article. To strengthen the argument, in Fig. \ref{fig:appB} are shown three additional examples, showcasing projection forms observed in both the classical trajectories with critical energy $E_{\mathrm{Classic}}$ and in the quantum Husimi functions $\mathfrak{H}_{E}^{\mathcal{T}}(N_1,N_3)$ and $\mathfrak{H}_{E}(N_1,N_3)$, for eigenstates with eigenvalues $E \approx E_{\mathrm{Classic}}$. Identification is facilitated through visual comparison.


\newpage
\bibliographystyle{ieeetr}
\bibliography{biblioCheck2}


\end{document}